\documentclass[12pt]{revtex4}
\usepackage{graphics,bm}
\usepackage{amssymb}
\usepackage{epsfig}
\usepackage{epsf}
\def\frac#1#2{{\textstyle{#1 \over #2}}}

\def\Im{{\rm Im}}
\def\be{\begin{equation}} \def\ee{\end{equation}}
\def\beq{\begin{eqnarray}} \def\eeq{\end{eqnarray}}
\def\nn{\nonumber}

\def\vk{{\vec k}}

\begin{document}

\title{Recent Progress of Probing Correlated Electron States by Point Contact Spectroscopy}
\author{Wei-Cheng Lee}
\email{wlee@binghamton.edu}
\affiliation{Department of Physics, Applied Physics, and Astronomy, Binghamton University - State University of New York, Binghamton, USA}
\author{Laura H. Greene}
\email{lhgreene@magnet.fsu.edu}
\affiliation{National High Magnetic Field Laboratory and Department of Physics, Florida State University, Tallahassee, Florida 32306, USA}

\date{\today}

\begin{abstract}
We review recent progress in point contact spectroscopy (PCS) to extract spectroscopic information out of correlated electron materials, with the emphasis on non-superconducting states. 
PCS has been used to detect bosonic excitations in normal metals, where signatures (e.g. phonons) are usually less than 1$\%$ of the measured conductance.
In the superconducting state, point contact Andreev reflection (PCAR) has been widely used to study properties of the superconducting gap in various superconductors. It has been 
well-recognized that the corresponding conductance can be accurately fitted by the Blonder-Tinkham-Klapwijk (BTK) theory in which the AR occurring near the point contact junction
is modeled by three parameters; the superconducting gap, the quasiparticle scattering rate, and a dimensionless parameter, $Z$, 
describing the strength of the potential barrier at the junction. 
AR can be as large as 100$\%$ of the background conductance, and only arises in the case of superconductors.
In the last decade, there have been more and more 
experimental results suggesting that the point contact conductance could reveal new features associated with the unusual single electron dynamics in non-superconducting states, 
shedding a new light on exploring the nature of the competing phases in correlated materials. To correctly interpret these new features, it is crucial to re-examine the modeling of 
the point contact junctions, the formalism used to describe the single electron dynamics particularly in point contact spectroscopy, and the physical quantity that should be computed to 
understand the conductance. We will summarize the theories for point contact spectroscopy developed from different approaches and highlight these conceptual differences distinguishing
point contact spectroscopy from tunneling-based probes. Moreover, we will show how the Schwinger-Kadanoff-Baym-Keldysh (SKBK) formalism together with the appropriate modeling of the 
nano-scale point contacts randomly distributed across the junction leads to the conclusion that the point contact conductance is proportional to the {\it effective density of states}, a physical 
quantity that can be computed if the electron self energy is known. The experimental data on iron based superconductors and heavy fermion compounds will be analyzed in this 
framework. These recent developments have extended the applicability of point contact spectroscopy to correlated materials, which will help us achieve a deeper understanding of the 
single electron dynamics in strongly correlated systems.

\end{abstract}

\maketitle

\section{Introduction}
Correlated electron materials have hosted numerous platforms for exotic quantum many body phenomena. Typically, conduction electrons in these materials occupy $d$ or $f$ levels which 
have narrow band widths, resulting 
in strong electron-electron correlations. These correlations give rise to a number of intriguing physical properties that cannot be captured 
by Fermi liquid theory, the conventional wisdom to understand the physics of simple metals.
Examples include, but are not limited to: high-temperature superconductivity\cite{dagottormp1994,armitagermp2010,stewartrmp2011,scalapinormp2012};
Mott insulator\cite{imadarmp1998,leermp2006}; heavy fermion compounds\cite{tsunetsugu1997,mydosh2011}; quantum nematic fluids \cite{fradkin2010}; and orbital ordering\cite{salamonrmp2001,leewcooreview}.
We wish to point out an important reason to study non-superconducting states. 
Unconventional superconductivity is defined by the superconducting order parameter having a lower symmetry than the underlying lattice (e.g., $d$-wave in the cuprates). 
The community has,
however, been calling the iron-based superconductors “unconventional” even their order parameter is $S_{\pm}$, which has the same symmetry as the underlying lattice. 
After an intensive research on these high-temperature superconductors, it has been widely believed that the normal state properties hold the key to understand the mechanism of the superconductivity,
and the gap symmetry is mainly a consequence of fluctuations which already exist in the normal states.
In addition, a frequent feature among various unconventional superconductors is the existence of a region in the phase diagram where superconductivity is absent 
and the system exhibits physical properties that cannot be captured by Fermi liquid theory. 
Consequently, the investigation on non-superconducting states is critically important for this type of unconventional superconductor, which is the main scope of this review.

For the study of correlated materials, spectroscopic measurements are particularly valuable because they could construct a complete atlas of the electronic dynamics as functions of the frequency, the 
temperature, and sometimes the momentum. Roughly speaking, these spectroscopic measurements may be categorized into two groups depending on their associated correlation functions.
For example, the inelastic neutron scattering measures the {\it spin-spin} correlation function\cite{kivelson2003,dai2015}, DC and optical conductivities measure
the {\it current-current} correlation function\cite{mahan}, 
and the electron energy loss spectroscopy (EELS) measures the {\it density-density} correlation function.\cite{kogar2014}
These correlation functions are {\it two-particle} correlation functions, meaning that they describe the dynamics involving scatterings between pairs of particles.
In a non-interacting system, these pairs do not interact with each other, leading to a broad spectrum in the corresponding correlation functions known as the particle-hole continuum. 
The presence of interactions
may result in sharp peaks in the correlation functions. These peaks are associated with certain collective excitations out of all electrons in the system which do not exist in the single 
particle 
dynamics. Consequently, valuable information about the interaction effects can be extracted from the collective excitations observed in these measurements.

While the collective excitations are certainly important, they contain limited information about the single electron dynamics. The reason is that two particle correlation functions are often
subject to not only the scatterings of electrons, but also some conservation laws. 
The best example is the conservation of the total momentum of the electrons that must be taken into account, both in the current-current and 
the density-density correlation functions\cite{turlakov2003,leewcmir2015}.
If the system contains only the Coulomb interaction which conserves the total momentum of electrons during the scattering process, the conductivity would always be infinite even though
the single electron dynamics could change dramastically as the strength of the Coulomb interaction varies\cite{kontani2008}.
Extensive research efforts have been spent on calculations of the single electron self-energy using non-perturbative approaches, and various hallmark features 
of correlated materials exhibiting in the single electron self-energy have been theoretically 
predicted\cite{hertz1976,millis1993,oganesyan2001,kotliar2006,lawler2006,leewc2012nfl,phillips2013,lo2013,vanacore2014}.
This is the place where probes sensitive to single electron dynamics can play a crucial role, and these probes include photoemission and angle-resolved photoemission spectroscopies
(PES and ARPES)\cite{damascelli2003}, scanning tunneling microscopy (STM)\cite{fischer2007}, planar tunneling spectroscopy, and point contact spectroscopy.
In this review, we focus on point contact and tunneling spectroscopies, highlighting their common features and differences.
The basic working principle of these probes is that electrons are injected into the system under study, and the elastic and inelastic scatterings of these 
injected electrons inside the system, which are closely related to single electron self energy, are revealed by measuring the junction conductance. 
It is remarkable that the different ways that the electrons are injected into the system result in very different junction conductances, which 
distinguishes point contact spectroscopy from tunneling based spectroscopies. To correctly interpret the conductance, special attention must be paid to the modeling of the junctions, 
the formalism used to describe the single electron dynamics particular to the probes we are studying, and the physical quantity that should be computed to understand the conductance.
This is exactly the scope of this review.

This review is organized as follows. Sec. II summarizes the theories for point contact spectroscpy, clarifying the difference between point contact and tunneling
spectroscopies. Sec. III surveys recent data of point contact conductance measured in several correlated materials, with a focus on the non-superconducting states showing unusual 
correlation effects. Finally, the summary and outlook will be given in Sec. IV.

\section{Theory}
\subsection{From planar tunneling to point contact}
\label{pcstheory}
Point contact spectroscopy is one of the powerful probes motivated by the study of the superconductivity. Soon after the birth of the BCS theory for the superconductivity in 1957\cite{bcs}, 
the existence of the superconducting gap, $\Delta$, a hallmark feature in BCS theory, was quickly confirmed in the measurements of the tunneling current between a normal metal and a superconductor.
Giaever {\it et al.} \cite{giaever1960,nicol1960} found that the tunneling conductance between two metals seperated by a thin insulating layer can be well described by 
\be
\frac{dI}{dV} = \frac{2\pi}{\hbar}\vert M\vert^2 \rho_f (eV),
\label{didvgia}
\ee
where $\vert M\vert^2$ represents the transmission probability of the tunneling electrons, $eV$ is the bias voltage, and $\rho_f$ is the density of states of the material 
into which the electron tunnels. 
If one or both metals become superconducting, Eq. \ref{didvgia} can fit the experimental data accurately by using the superconducting density of states. 
According to the BCS theory, the low-lying excitations in the superconducting ground state are linear combinations of electrons and holes due to the correlated nature of Cooper pairs, and the 
resulting density of states is
\beq
\rho_f(\vert E\vert) &=& 0, \vert E\vert < \Delta,\nn\\
&=& \frac{\rho_n(0) \vert E\vert}{\sqrt{E^2-\Delta^2}}, \vert E\vert > \Delta,
\label{bcsdos}
\eeq
where $\rho_n(0)$ is the density of states at Fermi energy in the normal state. Although this result offered a strong evidence of a non-zero $\Delta$, it required an assumption of the matrix 
element $\vert M\vert^2$ being not only energy-independent, but also unchanged when the system undergoes the superconducting phase transition. 
This assumption on $\vert M\vert^2$ was later justified by Bardeen\cite{bardeen1961} to be a consequence of the barrier potential due to the thin insulating tunneling barrier.
Since the potential in the barrier is high, the wave function inside the barrier is exponentially small for both normal and superconducting states. 
As a result, $\vert M\vert^2$, which is proportional to the wave function overlap in the barrier, is insensitive to the difference between the normal and the superconducting states.
This explanation was later supported by Harrison who performed a microscopic calculation on $\vert M\vert^2$ from the band theory\cite{harrison1961}. 
He found that for the simple metal, $\vert M\vert^2$ is 
proportional to the Fermi velocity which can be expressed as $v_f = \frac{1}{\hbar}(\frac{\partial E}{\partial k_x})$. Since the density of states can be expressed as 
$\rho_f = \frac{L}{\pi}(\frac{\partial E}{\partial k_x})^{-1}$, the tunneling conductance across two simple metals separated by an insulating layer is a constant, obeying Ohm's law.
For the tunneling conductance between a metal and a superconductor, one has to use $v_f$ from the normal state and $\rho_f$ from the superconducting state given in Eq. \ref{bcsdos}, 
which gives a tunneling conductance exhibiting the superconducting density of states, as observed by Giaever {\it et al.}

Harrison's theorem\cite{harrison1961} has an important implication. For the tunneling conductance to reveal the variation in the density of states, the system has to be in the superconducting state. 
This unusual constraint was further employed by McMillan and Rowell\cite{mcmillan1965} to map out the phonon properties. 
Measuring the tunneling conductance across a planar tunneling junction with leads (Pb), 
they observed that as the Pb was in superconducting state at 0.8K, the tunneling conductance showed some nonlinearities which disappeared as soon as the Pb was driven to the normal state. 
They further found that the second derivative on the voltage $V$ with respect to the current $I$
was directly proportional to the Eliashberg function,
\be
\frac{d^2V}{dI^2} \propto \alpha^2(\omega)F(\omega)
\ee
where $F(\omega)$ is the density of phonon modes contributing to the superconducting pairing mechanism, and $\alpha^2(\omega)$ is the strength of electron-phonon coupling.

\begin{figure}
\begin{center}
\includegraphics{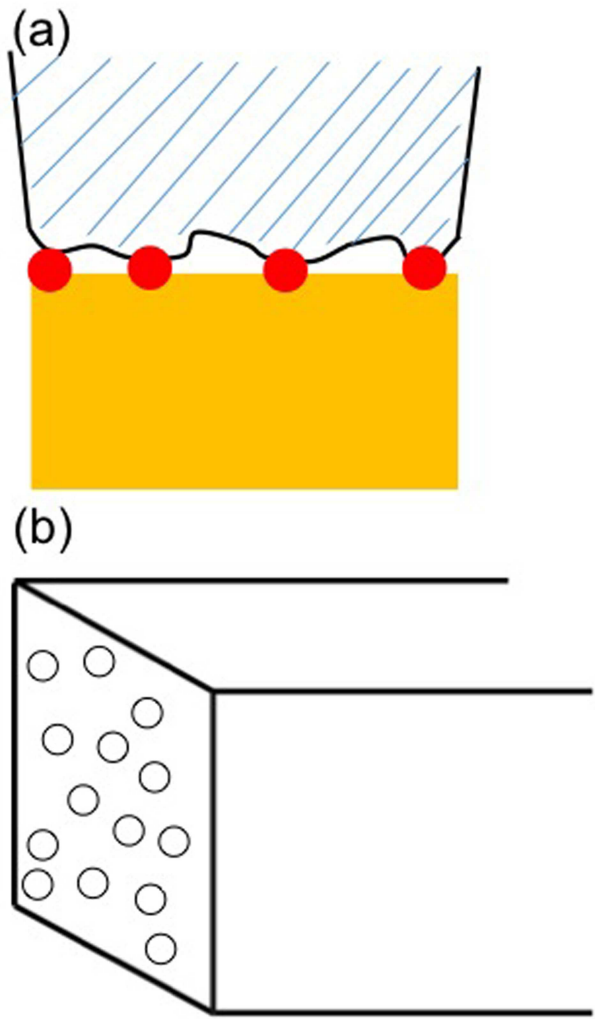}
\caption{\label{fig:realpcs} (a) Schematic illustration of the nano-scale metallic contacts. Electrons can be injected through these contacts (red dots) without tunneling processes. 
(b) The real-space profile of these randomly distributed point contacts.
These figures are adopted from Ref. [\onlinecite{leewcpnas}].}
\end{center}
\end{figure}

In 1974, while attemping to measure the tunneling conductance in a Pb/insulator/normal-metal planar junction, Yanson observed the nonlinearities in the tunneling conductance even 
when the Pb was driven to the normal state\cite{yanson1974}. He soon realized that these nonlinearities in the normal state 
were due to the junction being shorted with many nano-scale metallic contacts throughout the junction as shown in Fig. \ref{fig:realpcs}.
In this case, electrons are injected directly into the material without a tunneling process, and the assumption on $\vert M\vert^2$ used in Bardeen's and Harrison's theories is 
no longer valid. In other words, $\frac{dI}{dV}\propto \vert M\vert^2 \rho_f$ is not necessary a constant even in a non-superconducting state, if the junction contains these 
point-like nano-scale shorts. Yanson further showed that the ratio between the size of the point contacts ($d$) and the electron mean free path $l$ is critical. 
In the thermal regime ($d>>l$), the conductance becomes Ohmic without 
any spectroscopic information. In both the ballistic ($d<<l_{elastic}$) and the diffusive ($d<<l_{inelastic}$) regimes, spectroscopic information can be extracted from the conductance.
This pioneering work by Yanson marked the dawn of point contact spectroscopy\cite{khotkevich1995}.

It should be evident now that point contact spectroscopy is not a probe relying on a tunneling process. 
In the planar tunneling junction, the high barrier potential suppresses the electron wave function inside the insulating layer exponentially, resulting $\vert M\vert^2\propto v_f$ in both normal and 
superconducting states as explained by Bardeen and Harrison. In point contact spectroscopy,
the nano-scale metallic shorts on the junction permit electrons to be injected from one material to another directly without any tunneling processes, 
leading to a transmission probability $\vert M\vert^2$ {\it not} simply proportional to $v_f$. 
Moreover, in either ballistic or diffusive regimes, since electrons can enter the material without going through a potential barrier, 
they could possess excess energy to be relaxed by the inelastic scatterings inside the material. 
As a result, to theoretically study the point contact conductance, the modeling of the junction with point contacts and the inelastic scatterings inside the material must 
be taken into account correctly, and no concept based on the tunneling should be incorporated.
One of the most successful theories is the Blonder-Tinkham-Klapwijk (BTK) theory for the point contact conductance across the 
superconductor/insulator/normal-metal junction with point contacts\cite{btk1982}.

\subsection{Blonder-Tinkham-Klapwijk (BTK) theory}
While the planar tunneling of electrons from a normal metal to a superconductor gives zero conductance if $eV<\Delta$ as predicted by Eq. \ref{bcsdos}, 
the direct incidence of the electrons from a normal metal to a superconductor yields a 
totally different story. This is the case for a transparent metal/superconductor junction without an insulating layer between them. Due to the nature of the BCS pairing state, 
the incident electron can be reflected by the superconductor back to the normal metal as a hole, 
known as Andreev reflection\cite{andreev1964}.
As a result, the conductance across a transparent metal/superconductor junction may be as much as twice the normal state conductance at $eV<\Delta$ due to the Andreev reflection.
Then the question is: what should the point contact conductance across a metal/supercondcutor junction look like? Apparently, it must be composed of some tunneling, some normal and some 
Andreev reflections, and the correct theory must be able to 
account for results in these opposite limits of the planar tunneling and the transparent metal/superconductor junction, which is exactly the beauty of the BTK theory\cite{btk1982}.

To account for the effects from the interface, BTK theory starts from the Bogoliubov-de-Gennes (BdG) equations for superconductivity in the real space:
\beq
i\hbar\frac{\partial f(x,t)}{\partial t}&=& \big[-\frac{\hbar^2}{2m}\nabla_x^2 - \mu(x) + V(x)\big]f(x,t) + \Delta(x) g(x,t),\nn\\
i\hbar\frac{\partial g(x,t)}{\partial t}&=& -\big[-\frac{\hbar^2}{2m}\nabla_x^2 - \mu(x) + V(x)\big]g(x,t) + \Delta(x) f(x,t),
\label{bdg}
\eeq
where $f(x,t)$ and $g(x,t)$ are the wave functions for electron and hole respectively, $\mu(x)$ is the chemical potential, and $\Delta(x)$ is the gap function.
The crucial part of this model is the barrier potential $V(x) = Z\hbar v_F\delta(x)$, where $v_F$ is the Fermi velocity at the Fermi energy and $Z$ is a dimensionless parameter describing the 
strength of the barrier. If $Z$ is zero, it is corresponding to the case of a transparent metal/superconductor junction 
in which the complete Andreev reflection occurs at $eV<\Delta$. If $Z\geq 5$, it
corresponds to planar tunneling giving rise to zero conductance at $eV<\Delta$. 
Point contact spectroscopy usually corresponds to the case with a moderate value of $Z$ in which Andreev and the normal reflections coexist. As a result, in general point contact conductance has 
non-zero values at $eV<\Delta$. 

The BTK model has been known for its accurate fitting with the experimental data since it was developed in 1982. It has worked well not only for the conventional superconductors, but also for the 
unconventional supercondcutors emerging from correlated electron materials with two modifications. First, because the electrons suffer more scatterings near the interface, the 
quasiparticle lifetime, $\tau$, could be much shorter depending on the quality of the junction. To describe such an effect, the parameter $\Gamma = \frac{\hbar}{\tau}$ is introduced 
to fit the experimental data, and the BTK model becomes \cite{dynes1984,plecenik1994} 
\beq
i\hbar\frac{\partial f(x,t)}{\partial t}&=& \big[-\frac{\hbar^2}{2m}\nabla_x^2 - \mu(x) + V(x) - i\Gamma\big]f(x,t) + \Delta(x) g(x,t),\nn\\
i\hbar\frac{\partial g(x,t)}{\partial t}&=& -\big[-\frac{\hbar^2}{2m}\nabla_x^2 - \mu(x) + V(x) - i\Gamma\big]g(x,t) + \Delta(x) f(x,t).
\eeq
If the experimental data require a large value of $\Gamma$ as fitting, that would indicate a poor quality of the junction. 
Also, the BTK model is successfully used to simulate non s-wave\cite{tanaka1995,park2009,wu2010,daghero2011} and multiband superconductors\cite{brinkman2002}.
A complete simulation for $d$-wave superconductors with different values of $Z$ and $\Gamma$ can be found in Ref. [\onlinecite{arhamthesis}].

\subsection{Point contact conductance in correlated materials}
\label{pcscm}
Since point contact spectroscopy opens a new door to study non-superconducting, correlated states, it is intriguing to study the behavior of the point contact conductance in correlated materials.
In correlated electron materials, the strong electron-electron correlation significantly shortens the quasiparticle lifetime or may even destroy the quasiparticle picture, and these effects are, 
in principle, encoded in the electron self energy $\Sigma(\vec{k},\omega)$. Consequently, a formalism including the electron self energy would be necessary in order to correctly interpret the 
conductance. 

Motivated by recent experiments on iron based superconductors\cite{arham2012,arhamthesis}, Lee {\it et al.} have derived a formula to describe the point contact conductance in non-Fermi liquid
states based on the non-equilibrium Green function method known as Schwinger-Kadanoff-Baym-Keldysh (SKBK) formalism\cite{leewcpnas}.
Here we outline the working principle of point contact spectroscopy. With the presence of nano-scale contacts whose sizes are 
shorter than the electron mean-free path, the electron can be injected with excess energy through these contacts. 
This excess energy is relaxed only through the inelastic scatterings in the material, and all the 
excess energy should be exhausted as the electron reaches the electrode on the other side. The process described above contributes a measurable current, and the corresponding conductance $dI/dV$ 
depends on how the excess energy is relaxed throughout the material. In other words, the point contact conductance should reveal some information about the inelastic scatterings in the materials, 
and taking these irreversible relaxation processes into account is important for the theory of PCS conductance.

To see why the concept of the non-equilibrium is essential, we first review the equilibrium many body formalism\cite{mahan} as a comparison. It is assumed that at $t=-\infty$, the system is in
the simple ground state where all the interactions are absent. The interactions are turned on adiabatically from $t=-\infty$ to $t=0$ and then turned off adiabatically as $t\to \infty$. 
The equilibrium condition requires that the ground state at $t=\pm\infty$ must be the same up to a phase factor\cite{mahan}, and consequently the perturbation theory could be performed 
with respect to the same ground state at $t=\pm\infty$. In the setup of PCS, the ground state at $t=-\infty$ should be in the state where  the elctrodes and the 
system are completely decoupled. At zero bias voltage, we could assume that the electrodes and the system are already in thermally equilibrium states at $t=-\infty$, and the system would 
return to the same states at $t=\infty$. As a result, it is reasonable to impose the 
equilibrium condition if the bias voltage is zero. However, if the bias voltage is non-zero, there is no reason to assume that electrodes and the system are prepared in thermally 
equilibrium states at $t=-\infty$. 
After the interactions are fully turned on, 
the electrodes and the system will be re-thermalized via the contact Hamiltonian. As a result, as the interactions are turned off at $t=\infty$, the final states should in principle 
deviate from the initial states at $t=-\infty$, indicating the violation of the equilibrium condition. 
The SKBK formalism which allows the ground states at $t=\pm\infty$ to be different is an ideal approach to study this situation. In fact, 
the same argument has been made to study the spectroscopies involving tunneling processes as well\cite{keldysh}.

The next crucial issue is the contact Hamiltonian $H_c$. In the study of tunneling processes, a popular model which describes processes conserving momentum can be written as
\be
H_c = V_t \sum_{\vec{k},\sigma,\alpha\in L,R} d^\dagger_{\vec{k},\sigma,\alpha} c_{\vec{k},\sigma} + h.c.,\nn\\
\label{finalhc}
\ee
where $c^\dagger_{\vec{k},\sigma}$ creates an electron with momentum $\vec{k}$ and spin $\sigma$ in the system, and 
$d^\dagger_{\vec{k},\sigma,\alpha}$ is the creation operator for an electron in the right ($\alpha=R$) or left ($\alpha=L$)
electrode.
While Eq. \ref{finalhc} was originally proposed to describe the tunneling processes bewteen the electrodes and the system, it is quite intriguing that 
Eq. \ref{finalhc} can also be employed to describe the PCS conductance reasonaly well\cite{maltseva2009,fogelstr2010,fogelstr2014}. 
Lee {\it et al.} pointed out that with the consideration of the realistic point contact junctions, the contact Hamiltonian for PCS could be approximated as Eq. \ref{finalhc}. 
As a result, the usage of $H_c$ in Eq. \ref{finalhc} does not necessarily mean that the concept of the tunneling has to be adopted. Eq. \ref{finalhc} can represent the contact Hamiltonian for 
several types of spectroscopies depending on the strength of $V_t$. The argument is given below.
 
\begin{figure}
\includegraphics[width=3.14in]{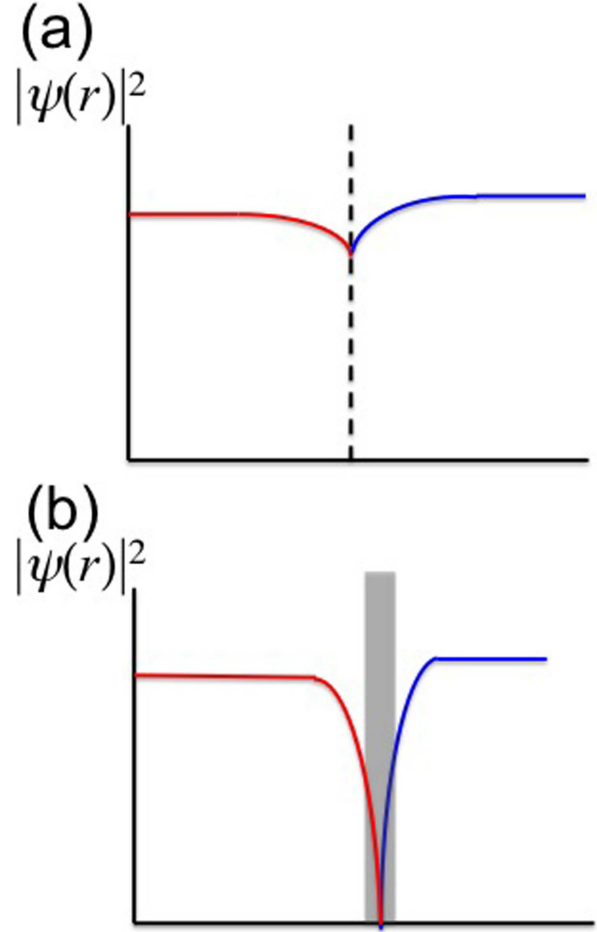}
\caption{\label{fig:waveoverlap} The profiles of the electron wave function $\vert\psi(r)\vert^2$ in (a) transparent and (b) planar tunneling junctions. 
(a) The dashed line refers to the interface, and the wave functions 
are finite on both sides. In this case, the transfer matrix element $V_t/\hbar v_F\sim 1$. (b) The dark area refers to the insulating layer in which the electron wave functions are expotentially 
suppressed, resulting in $V_t/\hbar v_F\to 0$.}
\end{figure}

The most general form of the contact Hamiltonian can be written in real space as
\be
H^g_c = \sum_{\sigma,\sigma'} \int d\vec{r} V^{\sigma,\sigma'}(\vec{r}) \psi^\dagger_{d,\sigma}(\vec{r}) \psi_{c,\sigma'}(\vec{r}) + h.c.,
\label{hc1}
\ee
where $\psi^\dagger_{d,c}(\vec{r})$ are the creation operators for
electrons in the electrode and the system. 
The form of $V^{\sigma,\sigma'}(\vec{r})$ is determined by the nature of the contact. 
For example, For the scanning tunneling microscopy (STM) in which electrons tunnel into the system via a nanoscale metallic tip, 
it is reasonable to set $V^{\sigma,\sigma'}(\vec{r}) = \delta_{\sigma,\sigma'}
V\delta(\vec{r})$. Its Fourier component is $V^{\sigma,\sigma'}_{\vec{k}.\alpha,\vec{p}} = \delta_{\sigma,\sigma'} \frac{V}{\Omega}$ for any pair of $(\vec{k},\vec{p})$, which
means that the electron can change its momentum as tunneling from the STM tip to the system.
In the opposite limit, for either the planar tunneling or the transparent junctions, $V^{\sigma,\sigma'}(\vec{r})$ can be approximated as a constant in the real space, and consequently
the corresponding Fourier component is
$V^{\sigma,\sigma'}_{\vec{k}.\alpha,\vec{p}} = \delta_{\sigma,\sigma'}V\delta(\vec{k}-\vec{p})$.
Therefore, the contact Hamiltonian in both cases can be reduced to Eq. \ref{finalhc}, 
though the former describes the tunneling process while the latter describes the process of the direct injection of the electrons.
Nevertheless, the strength of $V_t$ is quite different because of the wave function overlaps.
In the case of the planar tunneling, because there exists an insulating layer in the middle, the overlaps of the electron wave function in the contact region are very small. 
Therefore, Eq. \ref{finalhc} describes the case of the planar tunneling only in the limit of 
$V_t/\hbar v_F\to 0$, which corresponds to the limit of large $Z$ in BTK theory.
On the other hand, for the transparent junctions, since there is no insulating layer in between, the overlaps of the electron wave function in the contact region could be large, 
resulting in $V/\hbar v_F\sim 1$. 
This case is corresponding to the limit of $Z=0$ in BTK theory. The schematic demonstrations of the electron wave functions in the planar tunneling and the transparent junctions are 
plotted in Fig. \ref{fig:waveoverlap}.

Clearly, PCS should lie between two extreme cases of the STM and the planar tunneling/transparent junctions. 
On the one hand, because the size of the individual contact is at the nanoscale too, the processes without momentum conservation should be allowed. 
On the other hand, as shown in Fig. \ref{fig:realpcs}, a lot of the randomly distributed point contacts exist throughout the junction, which greatly suppresses the 
processes without momentum conservation.
Lee {\it et al.} argued that the 'classical' PCS is the one with many contacts, thus Eq. \ref{finalhc} is a good approximation with a moderate strength of 
$0<V_t/\hbar v_f << 1$. It is also corresponding to the case with low $Z$ in BTK theory.
The momentum dependence should be restored only as one is interested in 
the case of the quantum point contact spectroscopy, in which the size of the contact is small and the number of the contacts is low. 

The above discussion justifies that $H_c$ can be treated perturbatively with respect to $H_{electrode}$ and $H_{sys}$ for point contact spectroscopy. Such a perturbation theory has been developed 
by several authors, and we adopt the SKBK formalism given in Refs. [\onlinecite{keldysh}] and [\onlinecite{leewcpnas}] in which the conductance in the leading order of $V_t^2$ can be expressed as:
\be
\frac{dI}{dV} = \frac{e^2}{h} \int d\vec{p}_\parallel T(eV)\mathcal{A}(\vec{p}_\parallel,eV),
\label{finalg}
\ee
where $eV$ is the bias voltage, $\vec{p}_\parallel$ is the in-plane momentum,
\beq
\mathcal{A}(\vec{p}_\parallel,eV)&=&i\big[G^R(\vec{p}_\parallel,eV) - G^A(\vec{p}_\parallel,eV)\big]\nn\\
&=& \frac{{\rm Im}\Sigma(\vec{p}_\parallel,\epsilon)}{\big[eV - (E_{\vec{p}_\parallel}-\mu^\prime) -
{\rm Re}\Sigma(\vec{p}_\parallel,\epsilon)\big]^2 + \left[{\rm Im}\Sigma(\vec{p}_\parallel,\epsilon)\right]^2},\nn\\
\eeq
is the electron {\it spectral function}, and $G^{R,A}(\vec{p}_\parallel,eV)$ are the retarded and the advanced Green functions respectively with the self-energy including
contributions from scatterings inside the system as well as near the contacts with the electrodes,
\be
\Sigma(\vec{p}_\parallel,\epsilon) = \Sigma_{sys}(\vec{p}_\parallel,\epsilon) + \Sigma_{electrodes}(\vec{p}_\parallel,\epsilon).
\ee
$T(eV)$ is the transmission rate which depends on the details of the junction interface, and
$\mu^\prime = \mu - \frac{1}{2}eV$ contains the average chemical shift due to the electric field created by the bias voltage as discussed in Ref. [\onlinecite{keldysh}].

To see how Eq. \ref{finalg} is consistent with the other theories, we first consider the non-interacting system in which $\Sigma(\vec{p}_\parallel,\epsilon) = 0$. In this case, 
$\mathcal{A}(\vec{p}_\parallel,eV)$ reduces to $\delta(E-E_{\vec{p}_\parallel}+\mu^\prime)$ which is just the density of states at momentum $\vec{k}$. If $T(eV) \propto v_F$, Eq. \ref{finalg} 
recovers the constant conductance obtained by Harrison theorem for the simple metal. For the planar tunneling across the metal/insulator/superconductor junction, we keep $T(eV) \propto v_F$ and 
$\mathcal{A}(\vec{p}_\parallel,eV)$ in superconducting state leads to the density of states given in Eq. \ref{bcsdos}, which reproduces Bardeen's theory\cite{bardeen1961}. 
For point contact spectroscopy, $T(eV)$ is no longer proportional to $v_F$, and it could be approximated to be weakly energy-dependent for a good point contact junction.
As a result, the variation in the density of states due to the electron-phonon scatterings could be captured by setting the electron self energy to the one resulted from the electron-phonon scattering
\cite{mahan}, and the relation between $d^2 V/dI^2$ and the phonon density found by Yanson could be recovered.
For the BTK theory, since Andreev reflection is a phenomenon existing only at the interface, one has to solve the BdG equations in real space with $H_c$ included explicitly as shown in Eq. \ref{bdg}.
The resulting density of states will
contain the extra conductance at bias voltages smaller than the superconducting gap. 
As for the parameter $\Gamma$ introduced to represent the short quasiparticle lifetime near the interface, one can identify 
$\Gamma \sim \Im \Sigma_{electrodes}$ and this effect will be taken into account.

The advantage of Eq. \ref{finalg} is the direct inclusion of the electron self energy $\Sigma$ in the formalism. As soon as $\Sigma$ becomes non-zero, the spectral function 
$\mathcal{A}(\vec{p}_\parallel,eV)$ is no longer a simple Dirac delta function, leading to a significant redistribution of the electron spectral weight. 
Assuming $T(eV)\sim T$ for a good point contact junction, the conductance can now be rewritten as
\be
\frac{dI}{dV} \approx \frac{e^2 T}{h} \int d\vec{p}_\parallel \mathcal{A}(\vec{p}_\parallel,eV) = \frac{e^2 T}{h} D^{eff}(eV),
\label{finalg2}
\ee
where
\be
D^{eff}(E)\equiv \int d\vec{p}_\parallel \Im \mathcal{A}(\vec{p}_\parallel,eV)
\label{effectivedos}
\ee
is defined as the {\it effective density of states} if $\Sigma$ is non-zero, in contrast to the density of states for the non-interacting system 
$D(E)\equiv \int d\vec{p}_\parallel \delta(E-E_{\vec{p}_\parallel}+\mu^\prime)$. In other words, {\it the point contact conductance should be proportional to the effective density of states in correlated 
materials with non-zero electron self energy $\Sigma$. If the form of the electron self energy due to the electron-electron correlation is unusual, the point contact conductance should be able to 
reveal some information about it.} 

\subsection{Point contact conductance in the thermal regime}

\begin{figure}
\includegraphics{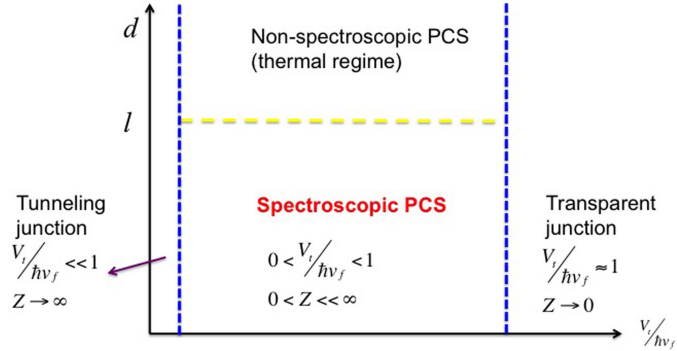}
\caption{\label{fig:dz-diagram} The applicability of the contact Hamiltonian given in Eq. \ref{finalhc} for different types of spectroscopy. $d$ is the contact size, $l$ is the electron 
mean-free path, $Z$ is the effective barrier parameter in BTK theory, and $V_t$ is the transfer parameter in Eq. \ref{finalhc}.}
\end{figure}

Lastly, we briefly comment about the point contact conductance in the thermal regime where the size of the contacts is bigger than the electron mean-free path. 
It has been a well-known fact that in the thermal regime, the point contact conductance loses spectroscopic features and strongly resembles the bulk conductivity\cite{naidyuk2005,park2009,park2012}.
As pointed out by Lee {\it et al.}, the reason is that in the thermal regime, the injected electrons have exhausted all the excess energy after passing through the contacts. As a 
result, the injected electrons are in thermal equilibrium with the rest of the electrons in the system immediately\cite{leewcpnas}.
Consequently, the conductance will be mainly dominated by the current induced by the electric field due to the bias voltage, which should be described by the current-current correlation function 
(Kubo formula) instead of the SKBK formalism describing the single electron dynamics. This explains the strong resemblance of the bulk conductivity to the point contact conductance in the thermal regime. 
In practice, checking this resemblance has been a useful way to judge the quality of the point contact junction\cite{park2009,park2012,arham2012}.
The applicability of the contact Hamiltonian given in Eq. \ref{finalhc} for different types of spectroscopy is summarized in Fig. \ref{fig:dz-diagram}.

\section{Experimental Data}

\subsection{Non-superconducting states in iron based superconductors}
Since the application of the point contact Andreev reflection spectroscopy to study the properties of superconducting gap in the iron based superconductors has been nicely reviewed recently 
\cite{daghero2011}, we will focus on the behavior of the point contact conductance in non-superconducting states in this review. 

\begin{figure}[t]
\includegraphics{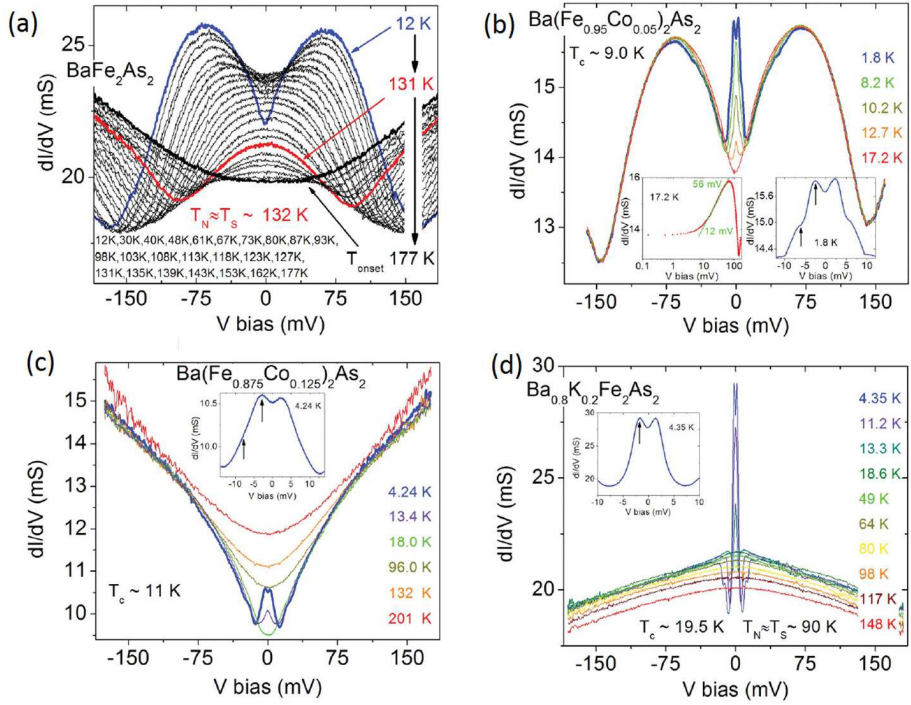}
\caption{These figures are taken from Ref.~\onlinecite{arham2012}. (a) Conductance spectra for $\rm{BaFe_2As_2}$. On top of a parabolic background, the conductance enhancement with peaks at $\sim$ 
65 meV was observed at low temperatures. As the temperature was increased, the double peaks merged into a single zero bias peak which survived well above $T_S$ (red curve). 
(b) In $\rm{Ba(Fe_{0.95}Co_{0.05})_2As_2}$, the superconductivity emerged in the presence of the long range magnetic order. At low temperatures, clear Andreev peaks were observed [right inset (b); 
the Andreev peaks are indicated by the arrows]. The Andreev spectra appeared together with the conductance enhancement with peaks at $\sim$ 65 meV which evolved with the temperature as it did for 
$\rm{BaFe_2As_2}$. This enhancement increased logarithmically near zero bias [left inset (b)]. (c) The overdoped compound $\rm{Ba(Fe_{0.875}Co_{0.125})_2As_2}$ showed clear Andreev spectra below $T_c$
as well. No significant change of the conductance peaks at higher bias values was observed, unlike the Co underdoped or the parent compounds. 
(d) The hole underdoped $\rm{Ba_{0.8}K_{0.2}Fe_2As_2}$ sample exhibited a coexistence of the superconductivity and the magnetism too. 
However, the conductance showed only the Andreev spectra below $T_c$, 
and no higher bias conductance enhancement was observed. This was in contrast to the data obtained from electron underdoped $\rm{Ba(Fe_{1-x}Co_x)_2As_2}$ [Fig.~\ref{fig:pcs}(b)].}
\label{fig:pcs}
\end{figure}

Arham {\it et al.}\cite{arham2012} measured the point contact conductance spectra on $\rm{Ba(Fe_{1-x}Co_x)_2As_2}$ for the entire phase diagram, and they found a quite universal behavior at dopings 
less than $6\%$.  The representative data are plotted in Fig.~\ref{fig:pcs}. 
Let's start with the undoped parent compound shown in Fig.~\ref{fig:pcs}(a).
At the temperature higher than 177 K, more than 40 K above the structual phase transition temperature $T_S$, the conductance showed a flat feature near the zero bias up to $\pm 40$ meV, obeying the 
Ohm's law. At the bias voltage larger (smaller) than $\pm 40$ meV, the conductance increased quadratically, which might be attributed to the nonlinear effects at higher bias voltages. 
At the temperature below 177K, 
a novel enhancement of the conductance near the zero bias was observed, and the enhancement became stronger as the temperature was lowered.  
No dramatic change occurred as $T_S$ was crossed (red curve). As the temperature was further lowered, a double peak feature was superimposed on the parabolic background, and the peak positions kept moving 
outward with the decrease of the temperature. At the lowest temperature (12 K, blue curve), a dip at zero bias appeared and two asymmetric conductance peaks emerged at $\sim$ 65 meV.

For the underdoped $\rm{Ba(Fe_{1-x}Co_x)_2As_2}$ with $x=0.05$ shown in Fig.~\ref{fig:pcs}(b), in addition to the structure and the magnetic phase transitions present already in the parent 
compound, the superconductivity emerged and coexisted with the long-range magnetic order at $T_c\sim$ 9K. The Andreev reflection at low bias voltages in the superconducting state was clearly observed, 
indicating a good point contact junction. However, the double-peak structure at $\sim \pm65$ meV was also observed, resembling the conductance of the parent compound. 
Above the superconducting transition temperature $T_c$, the zero bias peak due to the Andreev reflection completely disappeared, but the high bias conductance evolved just as it did for the 
parent compound $\rm{BaFe_2As_2}$. To be specific, as the temperature increased from $T_c$, the double-peak structure moved inward and merged into a single zero bias peak at a certain temperature below 
$T_S$. Then this zero bias peak disappeared at a temperature about 25 K higher than $T_S$. The right inset in Fig.~\ref{fig:pcs}(b) shows a zoom in of the Andreev reflection features while the left 
inset plots the conductance spectra above $T_c$ on a log plot.

The conductance of the overdoped $\rm{Ba(Fe_{1-x}Co_x)_2As_2}$  with $x=0.125$ and $T_c\sim$ 11K is summarized in Fig.~\ref{fig:pcs}(c). The Andreev reflection was still observed unambiguously 
in the superconducting state, but unlike the underdoped or the parent compounds, no intriguing evolution of the higher bias conductance as a function of the temperature were detected. 
The only feature observed above $T_c$ was the flattening of the parabolic background with the increase of the temperature.

Arham {\it et al.}\cite{arham2012} also reported the conductance spectra of the hole underdoped $\rm{Ba_{1-x}K_{x}Fe_2As_2}$ [Fig.~\ref{fig:pcs}(d)]. 
The sample has a coexistence of magnetism and superconductivity like its electron-doped cousin ($T_N$ = $T_S$ $\sim$ 90 K, $T_{c}$ $\sim$ 20 K), but the condutance showed very distinct behaviors. 
Below $T_c$ clear Andreev reflection was observed. Above $T_c$, Andreev reflection disappeared and the only feature observed was a downward facing background that remained unchanged with any further 
increase in temperature. This is remarkably different from the situation in the electron underdoped $\rm{Ba(Fe_{1-x}Co_x)_2As_2}$.

\begin{figure}[t]
\includegraphics{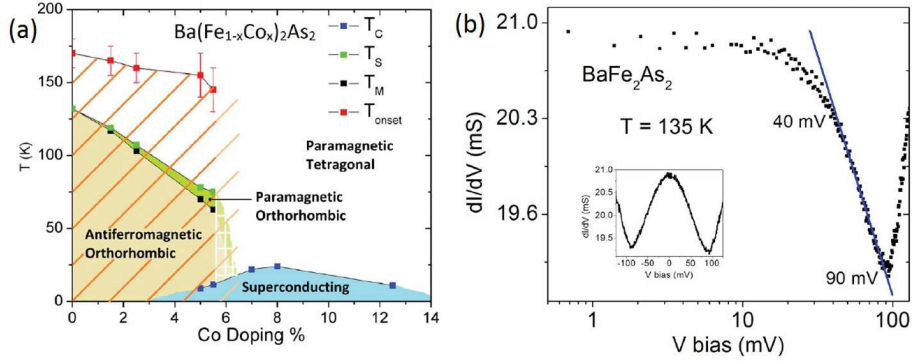}
\caption{Figures from Ref.~\onlinecite{arham2012}. (a) Revised phase diagram for $\rm{Ba(Fe_{1-x}Co_x)_2As_2}$ according to the conductance enhancement observed by point contact spectroscopy. 
(b) The logarithmic fitting of the conductance above $T_S$ for $\rm{BaFe_2As_2}$ from $\sim$ 40 meV to $\sim$ 90 meV.}
\label{fig:pcs-phase}
\end{figure}

A complete survey among different families of the iron based superconductors including $\rm{CaFe_2As_2}$, $\rm{SrFe_2As_2}$, and $\rm{Fe_{1.13}Te}$ leads to a conclusion that the novel zero bias peak 
appearing in the non-superconducting states is in close relation to the structural phase transition\cite{arham2012,arhamthesis}.
In particular, a correlation has been noticed between the presence of the conductance enhancement around zero bias and the in-plane resistivity anisotropy in the compounds. 
For the detwinned underdoped $\rm{AEFe_2As_2}$, it has been observed by various groups that while a resistivity anisotropy generally exists below $T_S$, 
the anisotropy in the resistivity above $T_S$ varies from materials to materials\cite{Chu2010,Tanatar,Fisher,Blomberg}.
Above $T_S$ there is a notable anisotropy for AE = Ba, a small anisotropy for AE = Sr, and no anisotropy for AE = Ca. 
The detwinned $\rm{Fe_{1+y}Te}$ also exhibits a resistivity anisotropy above the structural 
transition\cite{Jiang}. On the other hand, the detwinned underdoped $\rm{Ba_{1-x}K_{x}Fe_2As_2}$ does not show any anisotropy at all, either below or above $T_S$\cite{Ying}.
The appearance of the in-plane-resistivity anisotropy above $T_S$ in detwinned samples matches nicely with the existence of the conductance enhancement at zero bias in the non-superconducting state, 
strongly suggesting that they are likely resulted from the same underlying physics. 
A revised phase diagram for $\rm{Ba(Fe_{1-x}Co_x)_2As_2}$ shown in Fig.~\ref{fig:pcs-phase}(a) has been proposed by Arham {\it et al.}\cite{arham2012} to highlight the region where 
the zero bias conductance peak was observed, and it is evident that this region is strongly tied to the structure phase transition.

Lee {\it et al.}have proposed that orbital fluctuations above $T_S$ could be the key to understand the zero bias conductance observed in the non-superconducting state\cite{leewc2012nfl}. 
Their theoretical
interpretation is based on the orbital scenario in which the structure phase transition is identified as the second order transition of the orbital ordering in $d_{xz}$ and $d_{yz}$ orbitals. 
Furthermore, by analyzing the multi-orbital model involving the $d_{xz}$ and $d_{yz}$ orbitals, it is found that the orbital ordering transition is in fact a lattice version of the nematic transition 
due to $d$-wave Pomeranchuk instability\cite{leewc2009sr327}, and consequently the effects of the orbital fluctuations should be similar to those of the nematic fluctuations. As shown by Lawler 
{\it et al.} \cite{lawler2006}, the effective density of states modified by the nematic fluctuations defined in Eq. \ref{effectivedos} exhibited a logarithmic divergence at zero energy, and consequently 
the point contact conductance should have this divergence at zero bias according to Eq. \ref{finalg2}. 
Comparing the conductance data with the prediction given above, it has been found that the conductance enhancement for $\rm{BaFe_2As_2}$ above $T_S$ follows a log dependence from $\sim$ 40 meV to $\sim$ 90 meV
as shown in Fig.~\ref{fig:pcs-phase}(b). 
The discrepancy near the zero bias between the theory and the experimental data is due to the fact that the theoretical results are obtained in a Fermi system without lattice at the 
zero temperature.\cite{lawler2006} Lee {\it et al.} pointed out that both the lattice and the finite temperature effects would result in the flattening of the divergence near the zero bias.
It is remarkable that the data of $\rm{SrFe_2As_2}$ and $\rm{Fe_{1.13}Te}$ could be fitted in the same fashion. 
As for $\rm{Ba_{0.8}K_{0.2}Fe_2As_2}$, the conductance can not be fitted in the same way, suggesting that the orbital (nematic) fluctuations are much weaker in this material. This is also consistent 
with the absence of the resistivity anisotropy above $T_S$ in the detwinned samples. 
This analysis provides a strong evidence that the orbital (nematic) fluctuations play a central role in the enhancement of the zero-bias conductance in the non-superconducting state.

It should be noted that some details in the point contact conductance are still unexplained. First, the origin of the double peak structure deep inside the magnetic state is unclear. 
The naive interpretation would be the spin gap, but that interpretation could be falsified because the peak positions are very different in $\rm{Ba(Fe_{1-x}Co_x)_2As_2}$ and 
$\rm{Ba_{1-x}K_{x}Fe_2As_2}$ which have about the same size of spin gap. Second, the dip at zero bias appearing at lowest temperature in Fig. ~\ref{fig:pcs}(a) is not understood. 
The explanation for these details might require a realistic model to compute the energy-dependent $T(eV)$ in Eq. \ref{finalg}, which is beyond the scope of Eq. \ref{finalg2} neglecting the energy 
dependence of $T(eV)$. Further theoretical study will be necessary to settle down these problems.

\subsection{Heavy Fermion Systems}
\begin{figure}
\includegraphics[width=3.14in]{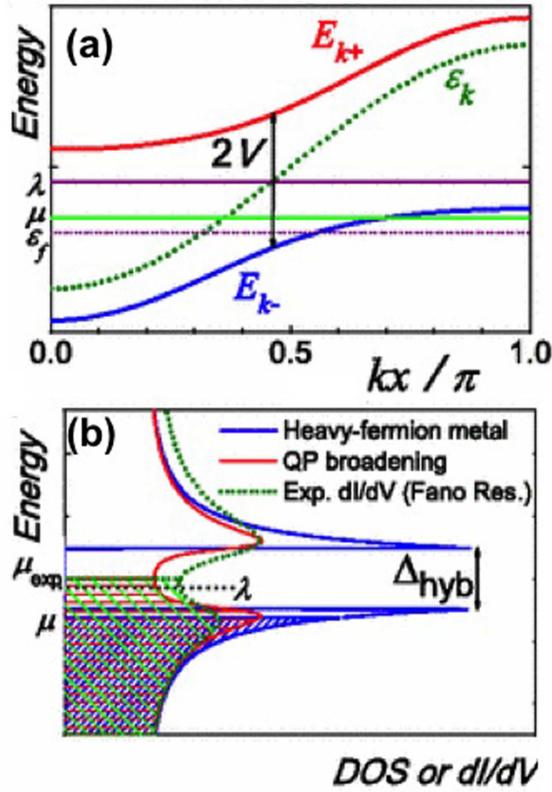}
\caption{\label{fig:hfband} (a) The mean-field band dispersions of the periodic Anderson model $H_{PAM}$ given in Eq. \ref{klek}. $\lambda$ is the renormalized band energy for $f$ electrons, 
$V=(1-n_f)^{1/2} V_0$, $V_0$ is the coupling between the conduction and $f$ electrons, $n_f$ is the average occupation number of $f$ electrons, and 
$\Delta_{hyb}\equiv 2V$.(b) The corresponding density of states (DOS) and the point contact conductance 
$dI/dV$. This figure is taken from Ref. [\onlinecite{park2012}].}
\end{figure}

The heavy fermion compounds are another type of correlated materials that have been studied intensively by point contact spectroscopy. These materials contain local magnetic moments from $f$-level
forming a lattice together with conduction electrons. The interplay between these local magnetic moments and the conduction electrons gives rise to a numer of many-body phenomena including 
Kondo effect, superconductivity, and quantum criticality. The simplest model describing the heavy fermion compounds is the periodic Anderson model, or its effective model obtained by the second 
order perturbation theory, the Kondo lattice model\cite{schrieffer1966,tsunetsugu1997}.
The Hamiltonian of the periodic Anderson model reads 
\beq
H_{PAM} &=& \sum_{\vk,\sigma} \epsilon(\vec{k}) c^\dagger_{\vk\sigma}c_{\vk\sigma} + \sum_i \epsilon_f f^\dagger_{i\sigma}f_{i\sigma}\nn\\
&+& V_0\sum_{\vk,\sigma} \big(f^\dagger_{\vk\sigma}c_{\vk\sigma} + H.c.\big) + U\sum_i n^f_{i\uparrow}n^f_{i\downarrow},
\eeq
where $c^\dagger$ and $f^\dagger$ are the creation operators for the conduction and the $f$ electrons, and $n^f_{i\sigma}=f^\dagger_{i\sigma} f_{i\sigma}$ is the number operator for 
$f$ electron on site $i$ with 
spin $\sigma$.
$H_{PAM}$ has been studied extensively. 
Below a critical temperature $T_c$, the mean-field theory obtains two renormalized bands with the hybridization between the conduction and the $f$ electrons, and the corresponding band 
dispersions are \cite{newns1987}
\beq
E_{\vk\pm} &=& \frac{1}{2}\big\{\epsilon(\vk) + \lambda \pm \sqrt{(\epsilon(\vk)-\lambda)^2 + 4V^2}\big\},\nn\\
V&=&(1-n_f)^{1/2} V_0,
\label{klek}
\eeq
where $\lambda$ is the renormalized band energy for $f$ electrons, and $n_f$ is the average occupation number of $f$ electrons.
It can be seen that $V$ naturally defines an energy scale for a band gap, namely the hybridization gap $\Delta_{hyb}$, and $\Delta_{hyb}$ should be reflected in the electronic density of states. 
In other words, point contact spectroscopy should be able to capture $\Delta_{hyb}$ according to the discussion in Sec. \ref{pcscm}, as shown in Fig. \ref{fig:hfband}.

\begin{figure}
\includegraphics[width=3.14in]{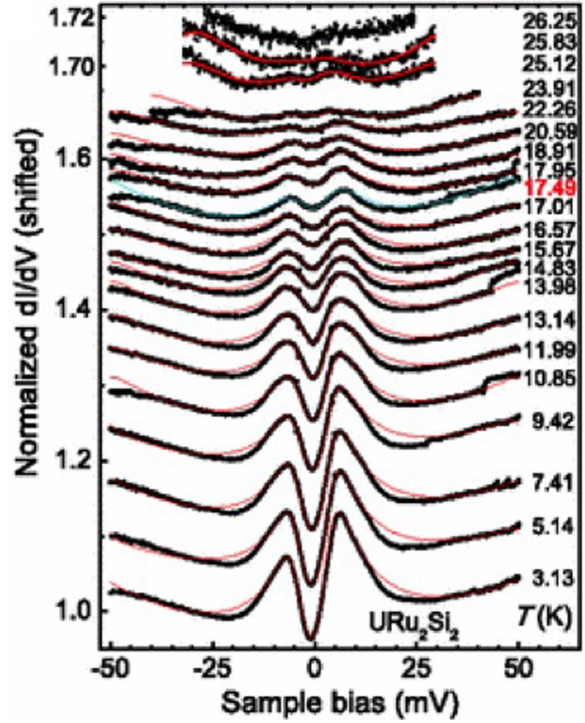}
\caption{\label{fig:pcs-122} Point contact couductance of URu$_2$Si$_2$ as a function of the temperature. The hybridization gap starts to emerge at temperatures well above 
 17.5 K, the critical temperature for the {\it hidden order}. This figure is taken from Ref. [\onlinecite{park2012}].}
\end{figure}

Let's first consider the case of the intermediate valence regime, i.e., $n_f \ll 1$. In this case, 
$\Delta_{hyb}$ should be sizable because of the factor $(1-n_f)^{1/2}$ in $V$ given in Eq. \ref{klek}, and the 
electronic density of states should have a significant suppression at the Fermi energy if the chemical potential lies in the hybridization gap. 
This should result in a dip accompanying with a double peak structure in the point 
contact conductance at zero bias according to the theory presented in Sec. \ref{pcscm}. Such a double peak structure has been clearly observed in the point contact conductance on URu$_2$Si$_2$ performed 
by Park {\it et al.}\cite{park2012}, as shown in Fig. \ref{fig:pcs-122}. By studying the evolution of the double peak structure with respect to the temperature, it has been concluded that the hybridization gap $\Delta_{hyb}$ appears 
well above 17.5 K, the critical temperature for a puzzling state known as {\it hidden order} whose origin remains mysteric
after intensive research efforts\cite{palstra1985,palstra1986,maple1986,mydosh2011}.
Despite the fact that the occurrence of certain phase transition has been indicated by a large suppression of the entropy observed at 17.5K, the corresponding order parameter is still under hot 
debate\cite{haule2009,elgazzar2009,oppeneer2010,dubi2011,chandra2013}.
Although the point contact conductance measured by Park {\it et al.} does not identify the correct order parameter for the hidden order in URu$_2$Si$_2$, it convincingly rules out theories using 
$\Delta_{hyb}$ as the order parameter. This work not only places a strong contraint on the theory for the hidden order in URu$_2$Si$_2$, but also demonstrates that point contact spectroscopy 
can be a powerful tool to detect the band renormalization driven by strong correlations.

\begin{figure}
\includegraphics[width=3.14in]{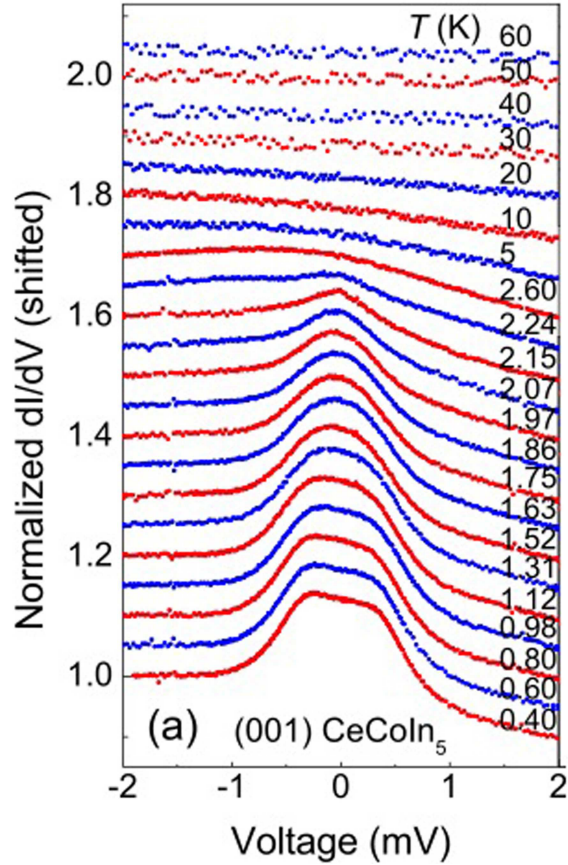}
\caption{\label{fig:pcs-115} Point contact couductance of CeCoIn$_5$ as a function of the temperature. The single peak around the zero bias with Fano lineshape starts to 
emerge at a temperature slightly above 
 2.3 K, the superconducting critical temperature. This figure is taken from Ref. [\onlinecite{park2008}].}
\end{figure}

In the Kondo regime, i.e., $n_f \approx 1$, $\Delta_{hyb}$ is vanishing and the double peak in the point contact conductance merges into a single peak near the zero bias, resembling the conductance 
from the Kondo impurity model\cite{maltseva2009,fogelstr2010}. Such a point contact conductance has been observed in Ce-based 115 materials Ce$M$In$_5$ for $M=$Co, Rh, and Ir
\cite{park2008,park2009,park20092}. Fig. \ref{fig:pcs-115} presents the point contact conductance of CeCoIn$_5$ as a function of the temperature. 
Recently, the heavy fermion compound YbAl$_3$ has been observed to exhibit a single peak near the zero bias in the point contact conductance, 
despite the fact that it has been identified as an intermediate valence compound by other measurements\cite{park2014}.
Nevertheless, a careful analysis on the small humps on top of the peak suggests that these humps are ressulted from the hybridization gap opening around 110 K, 
and the extra density of states filling the dip inside the gap should 
come from certain unknown correlation effects. More theoretical efforts will be required to address the sources of these correlation effects.

An interesting common feature observed in both intermediate ($n_f\ll 1$) and Kondo ($n_f\approx 1$) regimes is the Fano-like background in the conductance. The Fano line shape refers to the asymmetric 
profile of the conductance with respect to the zero bias, which was first discussed by Fano in 1961\cite{fano1961}.
In the Kondo-impurity system, the STM conductance due to the electron tunneling into the Kondo impurity also displays the Fano line shape, which has been theoretically understood as follows
\cite{maltseva2009,fogelstr2010,wolfle2010}.
The electrons from the STM tip have two tunneling channels, one to the conduction band and the other one to the impurity state. 
There exists a quantum interference between these two tunneling channels, which results in the asymmetric profile of the tunneling conductance. 
Since point contact spectroscopy does not rely on the electron tunneling as discussed in Sec. \ref{pcstheory}, one can not directly apply the same idea to explain the Fano line shape 
observed in the point contact conductance. Park {\it et al.} proposed that in point contact spectroscopy of the heavy fermion systems, the electrons can be injected into the 
two renormalized bands given in Eq. \ref{klek}, and the quantum interference between these two channels could cause the Fano line shape\cite{park20092}.
A later theory based on the SKBK formalism has shown that the asymmetry could be pronounced depending on the band renormalization\cite{fogelstr2010}.
It is a remarkable fact that the inclusion of an effective Fano line shape could usually lead to a much better fitting of the point contact conductance in the heavy fermion compounds
\cite{park2008,park2009,park20092,park2012,park2014}.

\section{Summary and Outlook}
In summary, the recent status of point contact spectroscopy in non-superconducting state has been reviewed. In the superconducting state, point contact spectroscopy has been used 
as one of the reliable tools to study the superconducting gap by virtue of BTK theory. In the normal metal, the nonlinearities in the conductance 
could be used to map out the phonon density. By a closer look into the working principle of point contact spectroscopy, it has been proposed that the point contact conductance is proportional to 
the {\it effective density of states} which is defined as the integration of the electron spectral function over the whole Brillouin zone, and this formalism is consistent with the previous 
theories. Moreover, it is emphasized that in point contact spectroscopy, electrons are injected directly via the nano-scale shorts randomly distributed throughout the juncion, and 
consequently no concept of the electron tunneling should be incorporated. 
The advantage of the formalism with the effective density of states becomes evident in correlated materials in which the electron 
self-energy could exhibit intriguing behaviors. By the definition of the effective density of states, the electron self energy is included automatically. We have shown that the point contact 
conductances in the heavy fermion compounds as well as in the non-superconducting state of the iron based superconductors could be understood within the same framework. These results 
have demonstrated that point contact spectroscopy could be a powerful tool to study the single electron dynamics of the strongly correlated systems.

The present theory still could not explain some details observed in the conductance. Particularly, the conductance near the zero bias at very low temperatures seems to indicate unknown 
correlation effects in the iron based superconductors and also in the heavy fermion compound YbAl$_3$. These details might be due to some elastic scatterings occuring near the junction which 
is not modeled properly in the present theory. The effects of these elastic scatterings near the junction might be negligible in weakly correlated metals but could be greatly enhanced in 
the strongly correlated systems. More research efforts will be necessary to resolve these issues.

\section{Acknowledgment}
We would like to thank H. Z. Arham, P. Aynajian, A. J. Leggett, P. W. Phillips, and W.-K. Park for helpful discussions.
This work is supported by the Center for Emergent Superconductivity, a DOE Energy Frontier Research Center, Grant No. DE- AC0298CH1088.


\begin{thebibliography}{77}
\expandafter\ifx\csname natexlab\endcsname\relax\def\natexlab#1{#1}\fi
\expandafter\ifx\csname bibnamefont\endcsname\relax
  \def\bibnamefont#1{#1}\fi
\expandafter\ifx\csname bibfnamefont\endcsname\relax
  \def\bibfnamefont#1{#1}\fi
\expandafter\ifx\csname citenamefont\endcsname\relax
  \def\citenamefont#1{#1}\fi
\expandafter\ifx\csname url\endcsname\relax
  \def\url#1{\texttt{#1}}\fi
\expandafter\ifx\csname urlprefix\endcsname\relax\def\urlprefix{URL }\fi
\providecommand{\bibinfo}[2]{#2}
\providecommand{\eprint}[2][]{\url{#2}}

\bibitem[{\citenamefont{Dagotto}(1994)}]{dagottormp1994}
\bibinfo{author}{\bibfnamefont{E.}~\bibnamefont{Dagotto}},
  \bibinfo{journal}{Rev. Mod. Phys.} \textbf{\bibinfo{volume}{66}},
  \bibinfo{pages}{763} (\bibinfo{year}{1994}).

\bibitem[{\citenamefont{Armitage et~al.}(2010)\citenamefont{Armitage, Fournier,
  and Greene}}]{armitagermp2010}
\bibinfo{author}{\bibfnamefont{N.~P.} \bibnamefont{Armitage}},
  \bibinfo{author}{\bibfnamefont{P.}~\bibnamefont{Fournier}}, \bibnamefont{and}
  \bibinfo{author}{\bibfnamefont{R.~L.} \bibnamefont{Greene}},
  \bibinfo{journal}{Rev. Mod. Phys.} \textbf{\bibinfo{volume}{82}},
  \bibinfo{pages}{2421} (\bibinfo{year}{2010}).

\bibitem[{\citenamefont{Stewart}(2011)}]{stewartrmp2011}
\bibinfo{author}{\bibfnamefont{G.~R.} \bibnamefont{Stewart}},
  \bibinfo{journal}{Rev. Mod. Phys.} \textbf{\bibinfo{volume}{83}},
  \bibinfo{pages}{1589} (\bibinfo{year}{2011}).

\bibitem[{\citenamefont{Scalapino}(2012)}]{scalapinormp2012}
\bibinfo{author}{\bibfnamefont{D.~J.} \bibnamefont{Scalapino}},
  \bibinfo{journal}{Rev. Mod. Phys.} \textbf{\bibinfo{volume}{84}},
  \bibinfo{pages}{1383} (\bibinfo{year}{2012}).

\bibitem[{\citenamefont{Imada et~al.}(1998)\citenamefont{Imada, Fujimori, and
  Tokura}}]{imadarmp1998}
\bibinfo{author}{\bibfnamefont{M.}~\bibnamefont{Imada}},
  \bibinfo{author}{\bibfnamefont{A.}~\bibnamefont{Fujimori}}, \bibnamefont{and}
  \bibinfo{author}{\bibfnamefont{Y.}~\bibnamefont{Tokura}},
  \bibinfo{journal}{Rev. Mod. Phys.} \textbf{\bibinfo{volume}{70}},
  \bibinfo{pages}{1039} (\bibinfo{year}{1998}).

\bibitem[{\citenamefont{Lee et~al.}(2006)\citenamefont{Lee, Nagaosa, and
  Wen}}]{leermp2006}
\bibinfo{author}{\bibfnamefont{P.~A.} \bibnamefont{Lee}},
  \bibinfo{author}{\bibfnamefont{N.}~\bibnamefont{Nagaosa}}, \bibnamefont{and}
  \bibinfo{author}{\bibfnamefont{X.-G.} \bibnamefont{Wen}},
  \bibinfo{journal}{Rev. Mod. Phys.} \textbf{\bibinfo{volume}{78}},
  \bibinfo{pages}{17} (\bibinfo{year}{2006}).

\bibitem[{\citenamefont{Tsunetsugu et~al.}(1997)\citenamefont{Tsunetsugu,
  Sigrist, and Ueda}}]{tsunetsugu1997}
\bibinfo{author}{\bibfnamefont{H.}~\bibnamefont{Tsunetsugu}},
  \bibinfo{author}{\bibfnamefont{M.}~\bibnamefont{Sigrist}}, \bibnamefont{and}
  \bibinfo{author}{\bibfnamefont{K.}~\bibnamefont{Ueda}},
  \bibinfo{journal}{Rev. Mod. Phys.} \textbf{\bibinfo{volume}{69}},
  \bibinfo{pages}{809} (\bibinfo{year}{1997}).

\bibitem[{\citenamefont{Mydosh and Oppeneer}(2011)}]{mydosh2011}
\bibinfo{author}{\bibfnamefont{J.~A.} \bibnamefont{Mydosh}} \bibnamefont{and}
  \bibinfo{author}{\bibfnamefont{P.~M.} \bibnamefont{Oppeneer}},
  \bibinfo{journal}{Rev. Mod. Phys.} \textbf{\bibinfo{volume}{83}},
  \bibinfo{pages}{1301} (\bibinfo{year}{2011}).

\bibitem[{\citenamefont{Fradkin et~al.}(2010)\citenamefont{Fradkin, Kivelson,
  Lawler, Eisenstein, and Mackenzie}}]{fradkin2010}
\bibinfo{author}{\bibfnamefont{E.}~\bibnamefont{Fradkin}},
  \bibinfo{author}{\bibfnamefont{S.~A.} \bibnamefont{Kivelson}},
  \bibinfo{author}{\bibfnamefont{M.~J.} \bibnamefont{Lawler}},
  \bibinfo{author}{\bibfnamefont{J.~P.} \bibnamefont{Eisenstein}},
  \bibnamefont{and} \bibinfo{author}{\bibfnamefont{A.~P.}
  \bibnamefont{Mackenzie}}, \bibinfo{journal}{Annual Review of Condensed Matter
  Physics} \textbf{\bibinfo{volume}{1}}, \bibinfo{pages}{153}
  (\bibinfo{year}{2010}).

\bibitem[{\citenamefont{Salamon and Jaime}(2001)}]{salamonrmp2001}
\bibinfo{author}{\bibfnamefont{M.~B.} \bibnamefont{Salamon}} \bibnamefont{and}
  \bibinfo{author}{\bibfnamefont{M.}~\bibnamefont{Jaime}},
  \bibinfo{journal}{Rev. Mod. Phys.} \textbf{\bibinfo{volume}{73}},
  \bibinfo{pages}{583} (\bibinfo{year}{2001}).

\bibitem[{\citenamefont{Lee et~al.}(2013)\citenamefont{Lee, Lv, and
  Arham}}]{leewcooreview}
\bibinfo{author}{\bibfnamefont{W.-C.} \bibnamefont{Lee}},
  \bibinfo{author}{\bibfnamefont{W.}~\bibnamefont{Lv}}, \bibnamefont{and}
  \bibinfo{author}{\bibfnamefont{H.~Z.} \bibnamefont{Arham}},
  \bibinfo{journal}{International Journal of Modern Physics B}
  \textbf{\bibinfo{volume}{27}}, \bibinfo{pages}{1330014}
  (\bibinfo{year}{2013}).

\bibitem[{\citenamefont{Kivelson et~al.}(2003)\citenamefont{Kivelson, Bindloss,
  Fradkin, Oganesyan, Tranquada, Kapitulnik, and Howald}}]{kivelson2003}
\bibinfo{author}{\bibfnamefont{S.~A.} \bibnamefont{Kivelson}},
  \bibinfo{author}{\bibfnamefont{I.~P.} \bibnamefont{Bindloss}},
  \bibinfo{author}{\bibfnamefont{E.}~\bibnamefont{Fradkin}},
  \bibinfo{author}{\bibfnamefont{V.}~\bibnamefont{Oganesyan}},
  \bibinfo{author}{\bibfnamefont{J.~M.} \bibnamefont{Tranquada}},
  \bibinfo{author}{\bibfnamefont{A.}~\bibnamefont{Kapitulnik}},
  \bibnamefont{and} \bibinfo{author}{\bibfnamefont{C.}~\bibnamefont{Howald}},
  \bibinfo{journal}{Rev. Mod. Phys.} \textbf{\bibinfo{volume}{75}},
  \bibinfo{pages}{1201} (\bibinfo{year}{2003}).

\bibitem[{\citenamefont{Dai}(2015)}]{dai2015}
\bibinfo{author}{\bibfnamefont{P.}~\bibnamefont{Dai}}, \bibinfo{journal}{Rev.
  Mod. Phys.} \textbf{\bibinfo{volume}{87}}, \bibinfo{pages}{855}
  (\bibinfo{year}{2015}).

\bibitem[{\citenamefont{Mahan}(2000)}]{mahan}
\bibinfo{author}{\bibfnamefont{G.~D.} \bibnamefont{Mahan}},
  \emph{\bibinfo{title}{Many-Particle Physics}} (\bibinfo{publisher}{Kluwer
  Academic / Plenum}, \bibinfo{year}{2000}).

\bibitem[{\citenamefont{Kogar et~al.}(2014)\citenamefont{Kogar, Vig, Gan, and
  Abbamonte}}]{kogar2014}
\bibinfo{author}{\bibfnamefont{A.}~\bibnamefont{Kogar}},
  \bibinfo{author}{\bibfnamefont{S.}~\bibnamefont{Vig}},
  \bibinfo{author}{\bibfnamefont{Y.}~\bibnamefont{Gan}}, \bibnamefont{and}
  \bibinfo{author}{\bibfnamefont{P.}~\bibnamefont{Abbamonte}},
  \bibinfo{journal}{Journal of Physics B: Atomic, Molecular and Optical
  Physics} \textbf{\bibinfo{volume}{47}}, \bibinfo{pages}{124034}
  (\bibinfo{year}{2014}).

\bibitem[{\citenamefont{Turlakov and Leggett}(2003)}]{turlakov2003}
\bibinfo{author}{\bibfnamefont{M.}~\bibnamefont{Turlakov}} \bibnamefont{and}
  \bibinfo{author}{\bibfnamefont{A.~J.} \bibnamefont{Leggett}},
  \bibinfo{journal}{Phys. Rev. B} \textbf{\bibinfo{volume}{67}},
  \bibinfo{pages}{094517} (\bibinfo{year}{2003}).

\bibitem[{\citenamefont{Lee}(2015)}]{leewcmir2015}
\bibinfo{author}{\bibfnamefont{W.-C.} \bibnamefont{Lee}},
  \bibinfo{journal}{Phys. Rev. B} \textbf{\bibinfo{volume}{91}},
  \bibinfo{pages}{224503} (\bibinfo{year}{2015}).

\bibitem[{\citenamefont{Kontani}(2008)}]{kontani2008}
\bibinfo{author}{\bibfnamefont{H.}~\bibnamefont{Kontani}},
  \bibinfo{journal}{Reports on Progress in Physics}
  \textbf{\bibinfo{volume}{71}}, \bibinfo{pages}{026501}
  (\bibinfo{year}{2008}).

\bibitem[{\citenamefont{Hertz}(1976)}]{hertz1976}
\bibinfo{author}{\bibfnamefont{J.~A.} \bibnamefont{Hertz}},
  \bibinfo{journal}{Phys. Rev. B} \textbf{\bibinfo{volume}{14}},
  \bibinfo{pages}{1165} (\bibinfo{year}{1976}).

\bibitem[{\citenamefont{Millis}(1993)}]{millis1993}
\bibinfo{author}{\bibfnamefont{A.~J.} \bibnamefont{Millis}},
  \bibinfo{journal}{Phys. Rev. B} \textbf{\bibinfo{volume}{48}},
  \bibinfo{pages}{7183} (\bibinfo{year}{1993}).

\bibitem[{\citenamefont{Oganesyan et~al.}(2001)\citenamefont{Oganesyan,
  Kivelson, and Fradkin}}]{oganesyan2001}
\bibinfo{author}{\bibfnamefont{V.}~\bibnamefont{Oganesyan}},
  \bibinfo{author}{\bibfnamefont{S.~A.} \bibnamefont{Kivelson}},
  \bibnamefont{and} \bibinfo{author}{\bibfnamefont{E.}~\bibnamefont{Fradkin}},
  \bibinfo{journal}{Phys. Rev. B} \textbf{\bibinfo{volume}{64}},
  \bibinfo{pages}{195109} (\bibinfo{year}{2001}).

\bibitem[{\citenamefont{Kotliar et~al.}(2006)\citenamefont{Kotliar, Savrasov,
  Haule, Oudovenko, Parcollet, and Marianetti}}]{kotliar2006}
\bibinfo{author}{\bibfnamefont{G.}~\bibnamefont{Kotliar}},
  \bibinfo{author}{\bibfnamefont{S.~Y.} \bibnamefont{Savrasov}},
  \bibinfo{author}{\bibfnamefont{K.}~\bibnamefont{Haule}},
  \bibinfo{author}{\bibfnamefont{V.~S.} \bibnamefont{Oudovenko}},
  \bibinfo{author}{\bibfnamefont{O.}~\bibnamefont{Parcollet}},
  \bibnamefont{and} \bibinfo{author}{\bibfnamefont{C.~A.}
  \bibnamefont{Marianetti}}, \bibinfo{journal}{Rev. Mod. Phys.}
  \textbf{\bibinfo{volume}{78}}, \bibinfo{pages}{865} (\bibinfo{year}{2006}).

\bibitem[{\citenamefont{Lawler et~al.}(2006)\citenamefont{Lawler, Barci,
  Fern\'andez, Fradkin, and Oxman}}]{lawler2006}
\bibinfo{author}{\bibfnamefont{M.~J.} \bibnamefont{Lawler}},
  \bibinfo{author}{\bibfnamefont{D.~G.} \bibnamefont{Barci}},
  \bibinfo{author}{\bibfnamefont{V.}~\bibnamefont{Fern\'andez}},
  \bibinfo{author}{\bibfnamefont{E.}~\bibnamefont{Fradkin}}, \bibnamefont{and}
  \bibinfo{author}{\bibfnamefont{L.}~\bibnamefont{Oxman}},
  \bibinfo{journal}{Phys. Rev. B} \textbf{\bibinfo{volume}{73}},
  \bibinfo{pages}{085101} (\bibinfo{year}{2006}).

\bibitem[{\citenamefont{Lee and Phillips}(2012)}]{leewc2012nfl}
\bibinfo{author}{\bibfnamefont{W.-C.} \bibnamefont{Lee}} \bibnamefont{and}
  \bibinfo{author}{\bibfnamefont{P.~W.} \bibnamefont{Phillips}},
  \bibinfo{journal}{Phys. Rev. B} \textbf{\bibinfo{volume}{86}},
  \bibinfo{pages}{245113} (\bibinfo{year}{2012}).

\bibitem[{\citenamefont{Phillips et~al.}(2013)\citenamefont{Phillips, Langley,
  and Hutasoit}}]{phillips2013}
\bibinfo{author}{\bibfnamefont{P.~W.} \bibnamefont{Phillips}},
  \bibinfo{author}{\bibfnamefont{B.~W.} \bibnamefont{Langley}},
  \bibnamefont{and} \bibinfo{author}{\bibfnamefont{J.~A.}
  \bibnamefont{Hutasoit}}, \bibinfo{journal}{Phys. Rev. B}
  \textbf{\bibinfo{volume}{88}}, \bibinfo{pages}{115129}
  (\bibinfo{year}{2013}).

\bibitem[{\citenamefont{Lo et~al.}(2013)\citenamefont{Lo, Hong, and
  Phillips}}]{lo2013}
\bibinfo{author}{\bibfnamefont{K.~W.} \bibnamefont{Lo}},
  \bibinfo{author}{\bibfnamefont{S.}~\bibnamefont{Hong}}, \bibnamefont{and}
  \bibinfo{author}{\bibfnamefont{P.~W.} \bibnamefont{Phillips}},
  \bibinfo{journal}{Phys. Rev. B} \textbf{\bibinfo{volume}{88}},
  \bibinfo{pages}{235114} (\bibinfo{year}{2013}).

\bibitem[{\citenamefont{Vanacore and Phillips}(2014)}]{vanacore2014}
\bibinfo{author}{\bibfnamefont{G.}~\bibnamefont{Vanacore}} \bibnamefont{and}
  \bibinfo{author}{\bibfnamefont{P.~W.} \bibnamefont{Phillips}},
  \bibinfo{journal}{Phys. Rev. D} \textbf{\bibinfo{volume}{90}},
  \bibinfo{pages}{044022} (\bibinfo{year}{2014}).

\bibitem[{\citenamefont{Damascelli et~al.}(2003)\citenamefont{Damascelli,
  Hussain, and Shen}}]{damascelli2003}
\bibinfo{author}{\bibfnamefont{A.}~\bibnamefont{Damascelli}},
  \bibinfo{author}{\bibfnamefont{Z.}~\bibnamefont{Hussain}}, \bibnamefont{and}
  \bibinfo{author}{\bibfnamefont{Z.-X.} \bibnamefont{Shen}},
  \bibinfo{journal}{Rev. Mod. Phys.} \textbf{\bibinfo{volume}{75}},
  \bibinfo{pages}{473} (\bibinfo{year}{2003}).

\bibitem[{\citenamefont{Fischer et~al.}(2007)\citenamefont{Fischer, Kugler,
  Maggio-Aprile, Berthod, and Renner}}]{fischer2007}
\bibinfo{author}{\bibfnamefont{O.}~\bibnamefont{Fischer}},
  \bibinfo{author}{\bibfnamefont{M.}~\bibnamefont{Kugler}},
  \bibinfo{author}{\bibfnamefont{I.}~\bibnamefont{Maggio-Aprile}},
  \bibinfo{author}{\bibfnamefont{C.}~\bibnamefont{Berthod}}, \bibnamefont{and}
  \bibinfo{author}{\bibfnamefont{C.}~\bibnamefont{Renner}},
  \bibinfo{journal}{Rev. Mod. Phys.} \textbf{\bibinfo{volume}{79}},
  \bibinfo{pages}{353} (\bibinfo{year}{2007}).

\bibitem[{\citenamefont{Bardeen et~al.}(1957)\citenamefont{Bardeen, Cooper, and
  Schrieffer}}]{bcs}
\bibinfo{author}{\bibfnamefont{J.}~\bibnamefont{Bardeen}},
  \bibinfo{author}{\bibfnamefont{L.~N.} \bibnamefont{Cooper}},
  \bibnamefont{and} \bibinfo{author}{\bibfnamefont{J.~R.}
  \bibnamefont{Schrieffer}}, \bibinfo{journal}{Phys. Rev.}
  \textbf{\bibinfo{volume}{108}}, \bibinfo{pages}{1175} (\bibinfo{year}{1957}).

\bibitem[{\citenamefont{Giaever}(1960)}]{giaever1960}
\bibinfo{author}{\bibfnamefont{I.}~\bibnamefont{Giaever}},
  \bibinfo{journal}{Phys. Rev. Lett.} \textbf{\bibinfo{volume}{5}},
  \bibinfo{pages}{147} (\bibinfo{year}{1960}).

\bibitem[{\citenamefont{Nicol et~al.}(1960)\citenamefont{Nicol, Shapiro, and
  Smith}}]{nicol1960}
\bibinfo{author}{\bibfnamefont{J.}~\bibnamefont{Nicol}},
  \bibinfo{author}{\bibfnamefont{S.}~\bibnamefont{Shapiro}}, \bibnamefont{and}
  \bibinfo{author}{\bibfnamefont{P.~H.} \bibnamefont{Smith}},
  \bibinfo{journal}{Phys. Rev. Lett.} \textbf{\bibinfo{volume}{5}},
  \bibinfo{pages}{461} (\bibinfo{year}{1960}).

\bibitem[{\citenamefont{Bardeen}(1961)}]{bardeen1961}
\bibinfo{author}{\bibfnamefont{J.}~\bibnamefont{Bardeen}},
  \bibinfo{journal}{Phys. Rev. Lett.} \textbf{\bibinfo{volume}{6}},
  \bibinfo{pages}{57} (\bibinfo{year}{1961}).

\bibitem[{\citenamefont{Harrison}(1961)}]{harrison1961}
\bibinfo{author}{\bibfnamefont{W.~A.} \bibnamefont{Harrison}},
  \bibinfo{journal}{Phys. Rev.} \textbf{\bibinfo{volume}{123}},
  \bibinfo{pages}{85} (\bibinfo{year}{1961}).

\bibitem[{\citenamefont{McMillan and Rowell}(1965)}]{mcmillan1965}
\bibinfo{author}{\bibfnamefont{W.~L.} \bibnamefont{McMillan}} \bibnamefont{and}
  \bibinfo{author}{\bibfnamefont{J.~M.} \bibnamefont{Rowell}},
  \bibinfo{journal}{Phys. Rev. Lett.} \textbf{\bibinfo{volume}{14}},
  \bibinfo{pages}{108} (\bibinfo{year}{1965}).

\bibitem[{\citenamefont{Yanson}(1974)}]{yanson1974}
\bibinfo{author}{\bibfnamefont{I.}~\bibnamefont{Yanson}}, \bibinfo{journal}{J.
  Exp. Theor. Phys.} \textbf{\bibinfo{volume}{39}}, \bibinfo{pages}{506}
  (\bibinfo{year}{1974}).

\bibitem[{\citenamefont{Khotkevich and Yanson}(1995)}]{khotkevich1995}
\bibinfo{author}{\bibfnamefont{A.~V.} \bibnamefont{Khotkevich}}
  \bibnamefont{and} \bibinfo{author}{\bibfnamefont{I.~K.}
  \bibnamefont{Yanson}}, \emph{\bibinfo{title}{Atlas of Point Contact Spectra
  of Electron-Phonon Interactions In Metals}} (\bibinfo{publisher}{Kluwer
  Academic}, \bibinfo{year}{1995}).

\bibitem[{\citenamefont{Lee et~al.}(2015)\citenamefont{Lee, Park, Arham,
  Greene, and Phillips}}]{leewcpnas}
\bibinfo{author}{\bibfnamefont{W.-C.} \bibnamefont{Lee}},
  \bibinfo{author}{\bibfnamefont{W.~K.} \bibnamefont{Park}},
  \bibinfo{author}{\bibfnamefont{H.~Z.} \bibnamefont{Arham}},
  \bibinfo{author}{\bibfnamefont{L.~H.} \bibnamefont{Greene}},
  \bibnamefont{and} \bibinfo{author}{\bibfnamefont{P.}~\bibnamefont{Phillips}},
  \bibinfo{journal}{Proc. Natl. Acad. Sci.} \textbf{\bibinfo{volume}{112}},
  \bibinfo{pages}{651} (\bibinfo{year}{2015}).

\bibitem[{\citenamefont{Blonder et~al.}(1982)\citenamefont{Blonder, Tinkham,
  and Klapwijk}}]{btk1982}
\bibinfo{author}{\bibfnamefont{G.~E.} \bibnamefont{Blonder}},
  \bibinfo{author}{\bibfnamefont{M.}~\bibnamefont{Tinkham}}, \bibnamefont{and}
  \bibinfo{author}{\bibfnamefont{T.~M.} \bibnamefont{Klapwijk}},
  \bibinfo{journal}{Phys. Rev. B} \textbf{\bibinfo{volume}{25}},
  \bibinfo{pages}{4515} (\bibinfo{year}{1982}).

\bibitem[{\citenamefont{Andreev}(1964)}]{andreev1964}
\bibinfo{author}{\bibfnamefont{A.~F.} \bibnamefont{Andreev}},
  \bibinfo{journal}{Sov. Phys.—JETP} \textbf{\bibinfo{volume}{19}},
  \bibinfo{pages}{1228} (\bibinfo{year}{1964}).

\bibitem[{\citenamefont{Dynes et~al.}(1984)\citenamefont{Dynes, Garno, Hertel,
  and Orlando}}]{dynes1984}
\bibinfo{author}{\bibfnamefont{R.~C.} \bibnamefont{Dynes}},
  \bibinfo{author}{\bibfnamefont{J.~P.} \bibnamefont{Garno}},
  \bibinfo{author}{\bibfnamefont{G.~B.} \bibnamefont{Hertel}},
  \bibnamefont{and} \bibinfo{author}{\bibfnamefont{T.~P.}
  \bibnamefont{Orlando}}, \bibinfo{journal}{Phys. Rev. Lett.}
  \textbf{\bibinfo{volume}{53}}, \bibinfo{pages}{2437} (\bibinfo{year}{1984}).

\bibitem[{\citenamefont{Plecen\'{\i}k et~al.}(1994)\citenamefont{Plecen\'{\i}k,
  Grajcar, Be\ifmmode \check{n}\else \v{n}\fi{}a\ifmmode~\check{c}\else
  \v{c}\fi{}ka, Seidel, and Pfuch}}]{plecenik1994}
\bibinfo{author}{\bibfnamefont{A.}~\bibnamefont{Plecen\'{\i}k}},
  \bibinfo{author}{\bibfnamefont{M.}~\bibnamefont{Grajcar}},
  \bibinfo{author}{\bibfnamefont{i.~c.~v.} \bibnamefont{Be\ifmmode
  \check{n}\else \v{n}\fi{}a\ifmmode~\check{c}\else \v{c}\fi{}ka}},
  \bibinfo{author}{\bibfnamefont{P.}~\bibnamefont{Seidel}}, \bibnamefont{and}
  \bibinfo{author}{\bibfnamefont{A.}~\bibnamefont{Pfuch}},
  \bibinfo{journal}{Phys. Rev. B} \textbf{\bibinfo{volume}{49}},
  \bibinfo{pages}{10016} (\bibinfo{year}{1994}).

\bibitem[{\citenamefont{Tanaka and Kashiwaya}(1995)}]{tanaka1995}
\bibinfo{author}{\bibfnamefont{Y.}~\bibnamefont{Tanaka}} \bibnamefont{and}
  \bibinfo{author}{\bibfnamefont{S.}~\bibnamefont{Kashiwaya}},
  \bibinfo{journal}{Phys. Rev. Lett.} \textbf{\bibinfo{volume}{74}},
  \bibinfo{pages}{3451} (\bibinfo{year}{1995}).

\bibitem[{\citenamefont{Park and Greene}(2009)}]{park2009}
\bibinfo{author}{\bibfnamefont{W.~K.} \bibnamefont{Park}} \bibnamefont{and}
  \bibinfo{author}{\bibfnamefont{L.~H.} \bibnamefont{Greene}},
  \bibinfo{journal}{Journal of Physics: Condensed Matter}
  \textbf{\bibinfo{volume}{21}}, \bibinfo{pages}{103203}
  (\bibinfo{year}{2009}).

\bibitem[{\citenamefont{Wu and Samokhin}(2010)}]{wu2010}
\bibinfo{author}{\bibfnamefont{S.}~\bibnamefont{Wu}} \bibnamefont{and}
  \bibinfo{author}{\bibfnamefont{K.~V.} \bibnamefont{Samokhin}},
  \bibinfo{journal}{Phys. Rev. B} \textbf{\bibinfo{volume}{81}},
  \bibinfo{pages}{214506} (\bibinfo{year}{2010}).

\bibitem[{\citenamefont{Daghero et~al.}(2011)\citenamefont{Daghero, Tortello,
  Ummarino, and Gonnelli}}]{daghero2011}
\bibinfo{author}{\bibfnamefont{D.}~\bibnamefont{Daghero}},
  \bibinfo{author}{\bibfnamefont{M.}~\bibnamefont{Tortello}},
  \bibinfo{author}{\bibfnamefont{G.}~\bibnamefont{Ummarino}}, \bibnamefont{and}
  \bibinfo{author}{\bibfnamefont{R.}~\bibnamefont{Gonnelli}},
  \bibinfo{journal}{Rep. Prog. Phys.} \textbf{\bibinfo{volume}{74}},
  \bibinfo{pages}{124509} (\bibinfo{year}{2011}).

\bibitem[{\citenamefont{Brinkman et~al.}(2002)\citenamefont{Brinkman, Golubov,
  Rogalla, Dolgov, Kortus, Kong, Jepsen, and Andersen}}]{brinkman2002}
\bibinfo{author}{\bibfnamefont{A.}~\bibnamefont{Brinkman}},
  \bibinfo{author}{\bibfnamefont{A.~A.} \bibnamefont{Golubov}},
  \bibinfo{author}{\bibfnamefont{H.}~\bibnamefont{Rogalla}},
  \bibinfo{author}{\bibfnamefont{O.~V.} \bibnamefont{Dolgov}},
  \bibinfo{author}{\bibfnamefont{J.}~\bibnamefont{Kortus}},
  \bibinfo{author}{\bibfnamefont{Y.}~\bibnamefont{Kong}},
  \bibinfo{author}{\bibfnamefont{O.}~\bibnamefont{Jepsen}}, \bibnamefont{and}
  \bibinfo{author}{\bibfnamefont{O.~K.} \bibnamefont{Andersen}},
  \bibinfo{journal}{Phys. Rev. B} \textbf{\bibinfo{volume}{65}},
  \bibinfo{pages}{180517} (\bibinfo{year}{2002}).

\bibitem[{\citenamefont{Arham}(2013)}]{arhamthesis}
\bibinfo{author}{\bibfnamefont{H.~Z.} \bibnamefont{Arham}}, Ph.D. thesis,
  \bibinfo{school}{University of Illinois at Urbana-Champaign}
  (\bibinfo{year}{2013}).

\bibitem[{\citenamefont{Arham et~al.}(2012)\citenamefont{Arham, Hunt, Park,
  Gillett, Das, Sebastian, Xu, Wen, Lin, Li et~al.}}]{arham2012}
\bibinfo{author}{\bibfnamefont{H.~Z.} \bibnamefont{Arham}},
  \bibinfo{author}{\bibfnamefont{C.~R.} \bibnamefont{Hunt}},
  \bibinfo{author}{\bibfnamefont{W.~K.} \bibnamefont{Park}},
  \bibinfo{author}{\bibfnamefont{J.}~\bibnamefont{Gillett}},
  \bibinfo{author}{\bibfnamefont{S.~D.} \bibnamefont{Das}},
  \bibinfo{author}{\bibfnamefont{S.~E.} \bibnamefont{Sebastian}},
  \bibinfo{author}{\bibfnamefont{Z.~J.} \bibnamefont{Xu}},
  \bibinfo{author}{\bibfnamefont{J.~S.} \bibnamefont{Wen}},
  \bibinfo{author}{\bibfnamefont{Z.~W.} \bibnamefont{Lin}},
  \bibinfo{author}{\bibfnamefont{Q.}~\bibnamefont{Li}}, \bibnamefont{et~al.},
  \bibinfo{journal}{Phys. Rev. B} \textbf{\bibinfo{volume}{85}},
  \bibinfo{pages}{214515} (\bibinfo{year}{2012}).

\bibitem[{\citenamefont{Haug and Jauho}(2007)}]{keldysh}
\bibinfo{author}{\bibfnamefont{H.~J.} \bibnamefont{Haug}} \bibnamefont{and}
  \bibinfo{author}{\bibfnamefont{A.-P.} \bibnamefont{Jauho}},
  \emph{\bibinfo{title}{Quantum Kinetics in Transport and Optics of
  Semiconductors}} (\bibinfo{publisher}{Springer}, \bibinfo{year}{2007}).

\bibitem[{\citenamefont{Maltseva et~al.}(2009)\citenamefont{Maltseva, Dzero,
  and Coleman}}]{maltseva2009}
\bibinfo{author}{\bibfnamefont{M.}~\bibnamefont{Maltseva}},
  \bibinfo{author}{\bibfnamefont{M.}~\bibnamefont{Dzero}}, \bibnamefont{and}
  \bibinfo{author}{\bibfnamefont{P.}~\bibnamefont{Coleman}},
  \bibinfo{journal}{Phys. Rev. Lett.} \textbf{\bibinfo{volume}{103}},
  \bibinfo{pages}{206402} (\bibinfo{year}{2009}).

\bibitem[{\citenamefont{Fogelstr\"om et~al.}(2010)\citenamefont{Fogelstr\"om,
  Park, Greene, Goll, and Graf}}]{fogelstr2010}
\bibinfo{author}{\bibfnamefont{M.}~\bibnamefont{Fogelstr\"om}},
  \bibinfo{author}{\bibfnamefont{W.~K.} \bibnamefont{Park}},
  \bibinfo{author}{\bibfnamefont{L.~H.} \bibnamefont{Greene}},
  \bibinfo{author}{\bibfnamefont{G.}~\bibnamefont{Goll}}, \bibnamefont{and}
  \bibinfo{author}{\bibfnamefont{M.~J.} \bibnamefont{Graf}},
  \bibinfo{journal}{Phys. Rev. B} \textbf{\bibinfo{volume}{82}},
  \bibinfo{pages}{014527} (\bibinfo{year}{2010}).

\bibitem[{\citenamefont{Fogelstr\"om et~al.}(2014)\citenamefont{Fogelstr\"om,
  Graf, Sidorov, Lu, Bauer, and Thompson}}]{fogelstr2014}
\bibinfo{author}{\bibfnamefont{M.}~\bibnamefont{Fogelstr\"om}},
  \bibinfo{author}{\bibfnamefont{M.~J.} \bibnamefont{Graf}},
  \bibinfo{author}{\bibfnamefont{V.~A.} \bibnamefont{Sidorov}},
  \bibinfo{author}{\bibfnamefont{X.}~\bibnamefont{Lu}},
  \bibinfo{author}{\bibfnamefont{E.~D.} \bibnamefont{Bauer}}, \bibnamefont{and}
  \bibinfo{author}{\bibfnamefont{J.~D.} \bibnamefont{Thompson}},
  \bibinfo{journal}{Phys. Rev. B} \textbf{\bibinfo{volume}{90}},
  \bibinfo{pages}{104512} (\bibinfo{year}{2014}).

\bibitem[{\citenamefont{Naidyuk and Yanson}(2005)}]{naidyuk2005}
\bibinfo{author}{\bibfnamefont{Y.}~\bibnamefont{Naidyuk}} \bibnamefont{and}
  \bibinfo{author}{\bibfnamefont{I.}~\bibnamefont{Yanson}},
  \emph{\bibinfo{title}{Point-Contact Spectroscopy}}, vol.
  \bibinfo{volume}{145} (\bibinfo{publisher}{Springer, New York},
  \bibinfo{year}{2005}).

\bibitem[{\citenamefont{Park et~al.}(2012)\citenamefont{Park, Tobash, Ronning,
  Bauer, Sarrao, Thompson, and Greene}}]{park2012}
\bibinfo{author}{\bibfnamefont{W.~K.} \bibnamefont{Park}},
  \bibinfo{author}{\bibfnamefont{P.~H.} \bibnamefont{Tobash}},
  \bibinfo{author}{\bibfnamefont{F.}~\bibnamefont{Ronning}},
  \bibinfo{author}{\bibfnamefont{E.~D.} \bibnamefont{Bauer}},
  \bibinfo{author}{\bibfnamefont{J.~L.} \bibnamefont{Sarrao}},
  \bibinfo{author}{\bibfnamefont{J.~D.} \bibnamefont{Thompson}},
  \bibnamefont{and} \bibinfo{author}{\bibfnamefont{L.~H.}
  \bibnamefont{Greene}}, \bibinfo{journal}{Phys. Rev. Lett.}
  \textbf{\bibinfo{volume}{108}}, \bibinfo{pages}{246403}
  (\bibinfo{year}{2012}).

\bibitem[{\citenamefont{Chu et~al.}(2010)\citenamefont{Chu, Analytis, De~Greve,
  McMahon, Islam, Yamamoto, and Fisher}}]{Chu2010}
\bibinfo{author}{\bibfnamefont{J.-H.} \bibnamefont{Chu}},
  \bibinfo{author}{\bibfnamefont{J.~G.} \bibnamefont{Analytis}},
  \bibinfo{author}{\bibfnamefont{K.}~\bibnamefont{De~Greve}},
  \bibinfo{author}{\bibfnamefont{P.~L.} \bibnamefont{McMahon}},
  \bibinfo{author}{\bibfnamefont{Z.}~\bibnamefont{Islam}},
  \bibinfo{author}{\bibfnamefont{Y.}~\bibnamefont{Yamamoto}}, \bibnamefont{and}
  \bibinfo{author}{\bibfnamefont{I.~R.} \bibnamefont{Fisher}},
  \bibinfo{journal}{Science} \textbf{\bibinfo{volume}{329}},
  \bibinfo{pages}{824} (\bibinfo{year}{2010}).

\bibitem[{\citenamefont{Tanatar et~al.}(2010)\citenamefont{Tanatar, Blomberg,
  Kreyssig, Kim, Ni, Thaler, Bud'ko, Canfield, Goldman, Mazin
  et~al.}}]{Tanatar}
\bibinfo{author}{\bibfnamefont{M.~A.} \bibnamefont{Tanatar}},
  \bibinfo{author}{\bibfnamefont{E.~C.} \bibnamefont{Blomberg}},
  \bibinfo{author}{\bibfnamefont{A.}~\bibnamefont{Kreyssig}},
  \bibinfo{author}{\bibfnamefont{M.~G.} \bibnamefont{Kim}},
  \bibinfo{author}{\bibfnamefont{N.}~\bibnamefont{Ni}},
  \bibinfo{author}{\bibfnamefont{A.}~\bibnamefont{Thaler}},
  \bibinfo{author}{\bibfnamefont{S.~L.} \bibnamefont{Bud'ko}},
  \bibinfo{author}{\bibfnamefont{P.~C.} \bibnamefont{Canfield}},
  \bibinfo{author}{\bibfnamefont{A.~I.} \bibnamefont{Goldman}},
  \bibinfo{author}{\bibfnamefont{I.~I.} \bibnamefont{Mazin}},
  \bibnamefont{et~al.}, \bibinfo{journal}{Phys. Rev. B}
  \textbf{\bibinfo{volume}{81}}, \bibinfo{pages}{184508}
  (\bibinfo{year}{2010}).

\bibitem[{\citenamefont{Fisher et~al.}(2011)\citenamefont{Fisher, Degiorgi, and
  Shen}}]{Fisher}
\bibinfo{author}{\bibfnamefont{I.~R.} \bibnamefont{Fisher}},
  \bibinfo{author}{\bibfnamefont{L.}~\bibnamefont{Degiorgi}}, \bibnamefont{and}
  \bibinfo{author}{\bibfnamefont{Z.~X.} \bibnamefont{Shen}},
  \bibinfo{journal}{Reports on Progress in Physics}
  \textbf{\bibinfo{volume}{74}}, \bibinfo{pages}{124506}
  (\bibinfo{year}{2011}).

\bibitem[{\citenamefont{Blomberg et~al.}(2011)\citenamefont{Blomberg, Tanatar,
  Kreyssig, Ni, Thaler, Hu, Bud'ko, Canfield, Goldman, and
  Prozorov}}]{Blomberg}
\bibinfo{author}{\bibfnamefont{E.~C.} \bibnamefont{Blomberg}},
  \bibinfo{author}{\bibfnamefont{M.~A.} \bibnamefont{Tanatar}},
  \bibinfo{author}{\bibfnamefont{A.}~\bibnamefont{Kreyssig}},
  \bibinfo{author}{\bibfnamefont{N.}~\bibnamefont{Ni}},
  \bibinfo{author}{\bibfnamefont{A.}~\bibnamefont{Thaler}},
  \bibinfo{author}{\bibfnamefont{R.}~\bibnamefont{Hu}},
  \bibinfo{author}{\bibfnamefont{S.~L.} \bibnamefont{Bud'ko}},
  \bibinfo{author}{\bibfnamefont{P.~C.} \bibnamefont{Canfield}},
  \bibinfo{author}{\bibfnamefont{A.~I.} \bibnamefont{Goldman}},
  \bibnamefont{and} \bibinfo{author}{\bibfnamefont{R.}~\bibnamefont{Prozorov}},
  \bibinfo{journal}{Phys. Rev. B} \textbf{\bibinfo{volume}{83}},
  \bibinfo{pages}{134505} (\bibinfo{year}{2011}).

\bibitem[{\citenamefont{Jiang et~al.}(2012)\citenamefont{Jiang, He, Zhang, Xu,
  Ge, Ye, Chen, Xie, and Feng}}]{Jiang}
\bibinfo{author}{\bibfnamefont{J.}~\bibnamefont{Jiang}},
  \bibinfo{author}{\bibfnamefont{C.}~\bibnamefont{He}},
  \bibinfo{author}{\bibfnamefont{Y.}~\bibnamefont{Zhang}},
  \bibinfo{author}{\bibfnamefont{M.}~\bibnamefont{Xu}},
  \bibinfo{author}{\bibfnamefont{Q.~Q.} \bibnamefont{Ge}},
  \bibinfo{author}{\bibfnamefont{Z.~R.} \bibnamefont{Ye}},
  \bibinfo{author}{\bibfnamefont{F.}~\bibnamefont{Chen}},
  \bibinfo{author}{\bibfnamefont{B.~P.} \bibnamefont{Xie}}, \bibnamefont{and}
  \bibinfo{author}{\bibfnamefont{D.~L.} \bibnamefont{Feng}}
  (\bibinfo{year}{2012}).

\bibitem[{\citenamefont{Ying et~al.}(2011)\citenamefont{Ying, Wang, Wu, Xiang,
  Liu, Yan, Wang, Zhang, Ye, Cheng et~al.}}]{Ying}
\bibinfo{author}{\bibfnamefont{J.~J.} \bibnamefont{Ying}},
  \bibinfo{author}{\bibfnamefont{X.~F.} \bibnamefont{Wang}},
  \bibinfo{author}{\bibfnamefont{T.}~\bibnamefont{Wu}},
  \bibinfo{author}{\bibfnamefont{Z.~J.} \bibnamefont{Xiang}},
  \bibinfo{author}{\bibfnamefont{R.~H.} \bibnamefont{Liu}},
  \bibinfo{author}{\bibfnamefont{Y.~J.} \bibnamefont{Yan}},
  \bibinfo{author}{\bibfnamefont{A.~F.} \bibnamefont{Wang}},
  \bibinfo{author}{\bibfnamefont{M.}~\bibnamefont{Zhang}},
  \bibinfo{author}{\bibfnamefont{G.~J.} \bibnamefont{Ye}},
  \bibinfo{author}{\bibfnamefont{P.}~\bibnamefont{Cheng}},
  \bibnamefont{et~al.}, \bibinfo{journal}{Phys. Rev. Lett.}
  \textbf{\bibinfo{volume}{107}}, \bibinfo{pages}{067001}
  (\bibinfo{year}{2011}).

\bibitem[{\citenamefont{Lee and Wu}(2009)}]{leewc2009sr327}
\bibinfo{author}{\bibfnamefont{W.-C.} \bibnamefont{Lee}} \bibnamefont{and}
  \bibinfo{author}{\bibfnamefont{C.}~\bibnamefont{Wu}}, \bibinfo{journal}{Phys.
  Rev. B} \textbf{\bibinfo{volume}{80}}, \bibinfo{pages}{104438}
  (\bibinfo{year}{2009}).

\bibitem[{\citenamefont{Schrieffer and Wolff}(1966)}]{schrieffer1966}
\bibinfo{author}{\bibfnamefont{J.~R.} \bibnamefont{Schrieffer}}
  \bibnamefont{and} \bibinfo{author}{\bibfnamefont{P.~A.} \bibnamefont{Wolff}},
  \bibinfo{journal}{Phys. Rev.} \textbf{\bibinfo{volume}{149}},
  \bibinfo{pages}{491} (\bibinfo{year}{1966}).

\bibitem[{\citenamefont{Newns and Read}(1987)}]{newns1987}
\bibinfo{author}{\bibfnamefont{D.}~\bibnamefont{Newns}} \bibnamefont{and}
  \bibinfo{author}{\bibfnamefont{N.}~\bibnamefont{Read}},
  \bibinfo{journal}{Advances in Physics} \textbf{\bibinfo{volume}{36}},
  \bibinfo{pages}{799} (\bibinfo{year}{1987}).

\bibitem[{\citenamefont{Palstra et~al.}(1985)\citenamefont{Palstra, Menovsky,
  Berg, Dirkmaat, Kes, Nieuwenhuys, and Mydosh}}]{palstra1985}
\bibinfo{author}{\bibfnamefont{T.~T.~M.} \bibnamefont{Palstra}},
  \bibinfo{author}{\bibfnamefont{A.~A.} \bibnamefont{Menovsky}},
  \bibinfo{author}{\bibfnamefont{J.~v.~d.} \bibnamefont{Berg}},
  \bibinfo{author}{\bibfnamefont{A.~J.} \bibnamefont{Dirkmaat}},
  \bibinfo{author}{\bibfnamefont{P.~H.} \bibnamefont{Kes}},
  \bibinfo{author}{\bibfnamefont{G.~J.} \bibnamefont{Nieuwenhuys}},
  \bibnamefont{and} \bibinfo{author}{\bibfnamefont{J.~A.}
  \bibnamefont{Mydosh}}, \bibinfo{journal}{Phys. Rev. Lett.}
  \textbf{\bibinfo{volume}{55}}, \bibinfo{pages}{2727} (\bibinfo{year}{1985}).

\bibitem[{\citenamefont{Palstra et~al.}(1986)\citenamefont{Palstra, Menovsky,
  and Mydosh}}]{palstra1986}
\bibinfo{author}{\bibfnamefont{T.~T.~M.} \bibnamefont{Palstra}},
  \bibinfo{author}{\bibfnamefont{A.~A.} \bibnamefont{Menovsky}},
  \bibnamefont{and} \bibinfo{author}{\bibfnamefont{J.~A.}
  \bibnamefont{Mydosh}}, \bibinfo{journal}{Phys. Rev. B}
  \textbf{\bibinfo{volume}{33}}, \bibinfo{pages}{6527} (\bibinfo{year}{1986}).

\bibitem[{\citenamefont{Maple et~al.}(1986)\citenamefont{Maple, Chen,
  Dalichaouch, Kohara, Rossel, Torikachvili, McElfresh, and
  Thompson}}]{maple1986}
\bibinfo{author}{\bibfnamefont{M.~B.} \bibnamefont{Maple}},
  \bibinfo{author}{\bibfnamefont{J.~W.} \bibnamefont{Chen}},
  \bibinfo{author}{\bibfnamefont{Y.}~\bibnamefont{Dalichaouch}},
  \bibinfo{author}{\bibfnamefont{T.}~\bibnamefont{Kohara}},
  \bibinfo{author}{\bibfnamefont{C.}~\bibnamefont{Rossel}},
  \bibinfo{author}{\bibfnamefont{M.~S.} \bibnamefont{Torikachvili}},
  \bibinfo{author}{\bibfnamefont{M.~W.} \bibnamefont{McElfresh}},
  \bibnamefont{and} \bibinfo{author}{\bibfnamefont{J.~D.}
  \bibnamefont{Thompson}}, \bibinfo{journal}{Phys. Rev. Lett.}
  \textbf{\bibinfo{volume}{56}}, \bibinfo{pages}{185} (\bibinfo{year}{1986}).

\bibitem[{\citenamefont{Haule and Kotliar}(2009)}]{haule2009}
\bibinfo{author}{\bibfnamefont{K.}~\bibnamefont{Haule}} \bibnamefont{and}
  \bibinfo{author}{\bibfnamefont{G.}~\bibnamefont{Kotliar}},
  \bibinfo{journal}{Nat. Phys.} \textbf{\bibinfo{volume}{5}},
  \bibinfo{pages}{796 } (\bibinfo{year}{2009}).

\bibitem[{\citenamefont{Elgazzar et~al.}(2009)\citenamefont{Elgazzar, Rusz,
  Amft, Oppeneer, and Mydosh}}]{elgazzar2009}
\bibinfo{author}{\bibfnamefont{S.}~\bibnamefont{Elgazzar}},
  \bibinfo{author}{\bibfnamefont{J.}~\bibnamefont{Rusz}},
  \bibinfo{author}{\bibfnamefont{M.}~\bibnamefont{Amft}},
  \bibinfo{author}{\bibfnamefont{P.~M.} \bibnamefont{Oppeneer}},
  \bibnamefont{and} \bibinfo{author}{\bibfnamefont{J.~A.}
  \bibnamefont{Mydosh}}, \bibinfo{journal}{Nat. Mater.}
  \textbf{\bibinfo{volume}{8}}, \bibinfo{pages}{337} (\bibinfo{year}{2009}).

\bibitem[{\citenamefont{Oppeneer et~al.}(2010)\citenamefont{Oppeneer, Rusz,
  Elgazzar, Suzuki, Durakiewicz, and Mydosh}}]{oppeneer2010}
\bibinfo{author}{\bibfnamefont{P.~M.} \bibnamefont{Oppeneer}},
  \bibinfo{author}{\bibfnamefont{J.}~\bibnamefont{Rusz}},
  \bibinfo{author}{\bibfnamefont{S.}~\bibnamefont{Elgazzar}},
  \bibinfo{author}{\bibfnamefont{M.-T.} \bibnamefont{Suzuki}},
  \bibinfo{author}{\bibfnamefont{T.}~\bibnamefont{Durakiewicz}},
  \bibnamefont{and} \bibinfo{author}{\bibfnamefont{J.~A.}
  \bibnamefont{Mydosh}}, \bibinfo{journal}{Phys. Rev. B}
  \textbf{\bibinfo{volume}{82}}, \bibinfo{pages}{205103}
  (\bibinfo{year}{2010}).

\bibitem[{\citenamefont{Dubi and Balatsky}(2011)}]{dubi2011}
\bibinfo{author}{\bibfnamefont{Y.}~\bibnamefont{Dubi}} \bibnamefont{and}
  \bibinfo{author}{\bibfnamefont{A.~V.} \bibnamefont{Balatsky}},
  \bibinfo{journal}{Phys. Rev. Lett.} \textbf{\bibinfo{volume}{106}},
  \bibinfo{pages}{086401} (\bibinfo{year}{2011}).

\bibitem[{\citenamefont{Chandra et~al.}(2013)\citenamefont{Chandra, Coleman,
  and Flint}}]{chandra2013}
\bibinfo{author}{\bibfnamefont{P.}~\bibnamefont{Chandra}},
  \bibinfo{author}{\bibfnamefont{P.}~\bibnamefont{Coleman}}, \bibnamefont{and}
  \bibinfo{author}{\bibfnamefont{R.}~\bibnamefont{Flint}},
  \bibinfo{journal}{Nature} \textbf{\bibinfo{volume}{493}},
  \bibinfo{pages}{621} (\bibinfo{year}{2013}).

\bibitem[{\citenamefont{Park et~al.}(2008)\citenamefont{Park, Sarrao, Thompson,
  and Greene}}]{park2008}
\bibinfo{author}{\bibfnamefont{W.~K.} \bibnamefont{Park}},
  \bibinfo{author}{\bibfnamefont{J.~L.} \bibnamefont{Sarrao}},
  \bibinfo{author}{\bibfnamefont{J.~D.} \bibnamefont{Thompson}},
  \bibnamefont{and} \bibinfo{author}{\bibfnamefont{L.~H.}
  \bibnamefont{Greene}}, \bibinfo{journal}{Phys. Rev. Lett.}
  \textbf{\bibinfo{volume}{100}}, \bibinfo{pages}{177001}
  (\bibinfo{year}{2008}).

\bibitem[{\citenamefont{Park et~al.}(2009)\citenamefont{Park, Bauer, Sarrao,
  Thompson, and Greene}}]{park20092}
\bibinfo{author}{\bibfnamefont{W.~K.} \bibnamefont{Park}},
  \bibinfo{author}{\bibfnamefont{E.~D.} \bibnamefont{Bauer}},
  \bibinfo{author}{\bibfnamefont{J.~L.} \bibnamefont{Sarrao}},
  \bibinfo{author}{\bibfnamefont{J.~D.} \bibnamefont{Thompson}},
  \bibnamefont{and} \bibinfo{author}{\bibfnamefont{L.~H.}
  \bibnamefont{Greene}}, \bibinfo{journal}{Journal of Physics: Conference
  Series} \textbf{\bibinfo{volume}{150}}, \bibinfo{pages}{052207}
  (\bibinfo{year}{2009}).

\bibitem[{\citenamefont{{Park} et~al.}(2014)\citenamefont{{Park},
  {Narasiwodeyar}, {Dwyer}, {Canfield}, and {Greene}}}]{park2014}
\bibinfo{author}{\bibfnamefont{W.~K.} \bibnamefont{{Park}}},
  \bibinfo{author}{\bibfnamefont{S.~M.} \bibnamefont{{Narasiwodeyar}}},
  \bibinfo{author}{\bibfnamefont{M.}~\bibnamefont{{Dwyer}}},
  \bibinfo{author}{\bibfnamefont{P.~C.} \bibnamefont{{Canfield}}},
  \bibnamefont{and} \bibinfo{author}{\bibfnamefont{L.~H.}
  \bibnamefont{{Greene}}}, \bibinfo{journal}{ArXiv e-prints}
  (\bibinfo{year}{2014}), \eprint{1411.7073}.

\bibitem[{\citenamefont{Fano}(1961)}]{fano1961}
\bibinfo{author}{\bibfnamefont{U.}~\bibnamefont{Fano}}, \bibinfo{journal}{Phys.
  Rev.} \textbf{\bibinfo{volume}{124}}, \bibinfo{pages}{1866}
  (\bibinfo{year}{1961}).

\bibitem[{\citenamefont{W\"olfle et~al.}(2010)\citenamefont{W\"olfle, Dubi, and
  Balatsky}}]{wolfle2010}
\bibinfo{author}{\bibfnamefont{P.}~\bibnamefont{W\"olfle}},
  \bibinfo{author}{\bibfnamefont{Y.}~\bibnamefont{Dubi}}, \bibnamefont{and}
  \bibinfo{author}{\bibfnamefont{A.~V.} \bibnamefont{Balatsky}},
  \bibinfo{journal}{Phys. Rev. Lett.} \textbf{\bibinfo{volume}{105}},
  \bibinfo{pages}{246401} (\bibinfo{year}{2010}).

\end{thebibliography}
\end{document}